\documentclass[printer]{aa} 
\usepackage{txfonts}
\usepackage{graphicx}

\begin{document}

\title{All quiet in the outer halo:\\ chemical abundances in the globular cluster Pal~3
  \thanks{This paper includes data gathered with the 6.5 meter Magellan Telescopes located at Las Campanas Observatory, Chile. 
  Some of the data presented herein were obtained at the W.M. Keck Observatory, which is operated as a scientific partnership among 
  the California Institute of Technology, the University of California and the National Aeronautics and Space Administration. 
  The Observatory was made possible by the generous financial support of the W.M. Keck Foundation.}
}

\author{Andreas Koch\inst{1} 
  \and Patrick C\^ot\'e\inst{2}   
  \and Andrew McWilliam\inst{3}
  }
  
\authorrunning{Koch, C\^ot\'e, \& McWilliam}
\titlerunning{Chemical abundances in the outer halo cluster Pal~3}
\offprints{A. Koch;  \email{ak326@astro.le.ac.uk}}

\institute{Department of Physics \& Astronomy, University of Leicester, University Road, Leicester 
LE1 7RH, UK
  \and National Research Council of Canada, Herzberg Institute of Astrophysics, 
  5071 West Saanich Road, Victoria, BC V9E 2E7, Canada 
  \and Carnegie Observatories, 813 Santa Barbara St., Pasadena, CA 91101, USA}

\date{}

\abstract {Globular clusters (GCs) in the outer halo are important probes of the composition and origin of the  Galactic stellar halo.}
 {We derive chemical element abundance ratios 
  in red giants belonging to the remote ($R\sim$90 kpc) GC Pal~3 and compare 
our  measurements to those for red giant stars in both inner and outer halo GCs.}
 {From high-resolution spectroscopy of four red giants, obtained with the Magellan/MIKE spectrograph at moderately high S/N,  
  we derive chemical abundances for 25 $\alpha$-, iron peak-, and neutron-capture elements. 
These abundance ratios are confirmed  by co-adding low S/N HIRES spectra of 19 stars along the red giant branch.} 
{Pal~3 shows $\alpha$-enhanced abundance patterns, and also its Fe-peak and neutron-capture element ratios, are fully compatible with those 
found in halo field stars and representative inner halo GCs of the same metallicity (such as M~13). The heavy elements in Pal 3 appear to be governed 
by $r$-process nucleosynthesis.  
Our limited sample does not show any significant star-to-star abundance variations in this cluster, although a weak Na-O anti-correlation cannot be ruled 
out by the present data.}
{Pal~3 thus appears as an archetypical GC with abundance ratios dissimilar to 
 dwarf spheroidal stars, ruling out a direct connection to such external systems. 
This conclusion is underscored by the lack of significant abundance spreads in this 
GC, in contrast to the broad abundance distributions seen in the dwarf galaxies.  Pal~3 appears to 
have evolved chemically coeval with the majority of GCs belonging to the Galactic inner $and$ outer halo, experiencing a similar enrichment history.}
\keywords{Stars: abundances -- Galaxy: abundances -- Galaxy: evolution -- Galaxy: halo -- globular clusters: individual: Pal~3}
\maketitle 
%
%
%
%
%
%
\section{Introduction}
As the oldest stellar systems in the universe, globular clusters (GCs) bear the imprints of the 
early formation and evolution epochs of the Milky Way (MW) system. In particular, the absence of   
a metallicity gradient in the outer halo\footnote{Many suggestions for the radius at which the outer halo separates from the inner halo are given 
in the literature, ranging from 8 to 30 kpc. The precise choice does not matter for Pal~3 since, at R$\sim$90 kpc, it is 
undoubtedly a member of the outer halo.}
led to the notion of an accretion origin for 
the  Galactic stellar halo that extended over several Gyr (Searle \& Zinn 1978). 
The separation into inner and outer halo populations has now been firmly established for both field stars and 
the GCs (e.g., Hartwick 1987; Norris \& Ryan 1989; Preston et al. 1991; Kinman et al. 1994; Carney et al. 1996; Chiba \& Beers 2000; Carollo et al. 2007; 
Lee et al. 2007; Miceli et al. 2008). 
This scenario is supported by the existence of a pronounced 
second-parameter problem among the outer halo GCs (Catelan et al. 2001) which  points to a broad age range within this population. 
Of prime importance are the chemical abundance patterns of halo field and GC stars (Freeman \& Bland-Hawthorne 2002; Pritzl et al. 2005). 
These are key observables that allow intercomparisons of the GCs to the 
dwarf spheroidal (dSph) galaxies (which are thought to have been accreted into the halo) and enable tests for (in)homogeneities among the inner 
and outer GC systems. 

Pal~3 is a faint ($M_V$ $\sim$ $-5.7$ mag) outer halo GC and, at a Galactocentric distance of $\sim$92 kpc (Stetson et al. 1999; Hilker 2006), 
one of only six known halo GCs at distances of $\sim$100 kpc or beyond.   
It is one of the most spatially-extended GCs, similar to the most compact ultrafaint dSph  (or most extended star cluster) 
candidates Willman~I (Willman et al. 2005) or Segue~I (Belokurov et al. 2007). At the same time, it is comparable in magnitude to  the 
ultrafaint Leo~IV (Belokurov et al. 2007) or Ursa Major~I (Zucker et al. 2006) systems, and therefore falls close to the gap between 
the dSphs and GCs in the magnitude-radius diagram (e.g., Gilmore et al. 2007). 
Lacking kinematic information, extended GCs like Pal~3 are therefore sometimes considered to be possible low-luminosity galaxies. 

Pal 3 is not part of any currently known stream (e.g., Palma et al. 2002) and it can be firmly excluded as member of the Sagittarius 
system (e.g, Bellazzini et al. 2003). Interestingly, within the large uncertainty of its proper motion,  its orbit is compatible both with being bound or unbound to the MW. 
Thus, it is conceivable that it has been captured by the Galaxy and is falling onto the MW for the first time (see also Chapman et al. 2007).  

Although the age estimates in the literature do not always agree (e.g., Stetson et al. 1999 vs. Vandenberg 2000), 
it seems clear that Pal~3 represents a halo GC, similar to the SMC cluster NGC~121 (Glatt et al. 2008),  
and that Pal~3 is likely 1--2 Gyr younger than inner halo GCs of the same metallicity such as M~3 and M~13 (e.g., Cohen \& Mel\'endez 2005a). 
An accurate age derivation, however, hinges on the assumption that ``they [Pal~3 and M3/M13] truly are chemically indistinguishable" (Vandenberg 2000).  
In this spirit, Cohen \& Mel\'endez (2005b) found that the outer halo GC NGC~7492 (R$_{\rm GC}$=25 kpc) has experienced a 
very similar enrichment history to the inner halo GCs, such as M3 or M13; 
in terms of their chemical abundances, these populations appear to be similar. 

All previous spectroscopic studies on Pal~3 have been carried out in low-dispersion mode (Ortolani \& Gratton 1989) and using the 
calcium triplet (CaT) metallicity indicator (Armandroff et al. 1992). Although no high-dispersion abundance study has been carried out for this remote cluster 
to place it in the context of the accretion scenario,  low-dispersion spectroscopy and colour magnitude diagram (CMD) studies have already 
established Pal~3 as a mildly metal poor system, with [Fe/H] estimates ranging from $-1.57$ to $-1.8$ dex
(Ortolani \& Gratton 1989; Armandroff et al. 1992; Stetson et al. 1999; Kraft \& Ivans 2003; Hilker 2006). 
In this paper, we aim to extend the chemical element information for objects in the Galactic halo out to larger distances,
and to present an initial characterisation of Pal~3's chemical abundance patterns. 
\section{Data}
\subsection{HIRES spectra}
During three nights in February and March 1999, we observed 25 stars in Pal 3 using the HIRES echelle spectrograph (Vogt et al. 1994) on the Keck I telescope.  
Our targets for this run were selected from a CMD constructed from $BV$ imaging obtained with the Low-Resolution Imaging Spectrometer (LRIS; Oke et al. 1995)
on the night of 13 January 1999. A CMD reaching roughly one magnitude below the main-sequence turnoff was constructed using short and long exposures
in both bandpasses (60s + 3$\times$180s in $V$, and 240s + 3$\times$420s in $B$). 

HIRES targets were identified from this CMD by selecting probable red 
giant branch (RGB) stars with $V \lesssim 20.25$.
Fig.~1 shows the location of these stars in the LRIS CMD. 
\begin{figure}[htb]
\centering
\includegraphics[width=1\hsize]{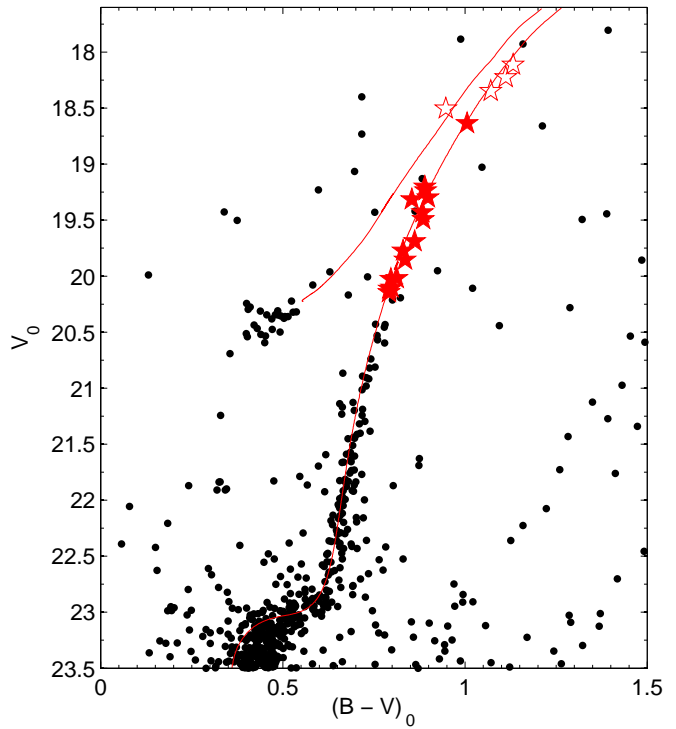}
\caption{Colour magnitude diagram of Pal~3 based on our LRIS photometry. 
Our HIRES (filled symbols) targets and those subsequently observed with MIKE (open symbols) are highlighted as red stars.
Also shown is an $\alpha$-enhanced Teramo isochrone (Pietrinferni et al. 2004) 
with an age of 10 Gyr and an [Fe/H] of $-$1.6 dex, shifted by our best reddening and distance modulus estimates of 0.04 and 19.80 mag. }
\end{figure}
We used a spectrograph setting that covers the wavelength range 
4450--6880\AA\ with spectral gaps between adjacent orders, 
a slit width of 1.15$\arcsec$ and a CCD binning of 2$\times$2 in the spatial and spectral directions. This gives
a spectral resolution of $R\approx34000$.   
 Each programme  star was observed for 420--2400~s, depending on its apparent magnitude (see Table~1). 
\begin{table}[htb]
\caption{Observing log}             
\centering          
\begin{tabular}{ccc}     
\hline\hline       
Object$^a$ & Date & Exposure time [s]\\
\hline
\multicolumn{3}{c}{HIRES} \\
\hline
Pal3-2  (H2)  & Feb 11 1999 & 1$\times$420  \\
Pal3-3  (H3)  & Feb 11 1999 & 1$\times$420  \\
Pal3-5  (---) & Mar 10 1999 & 1$\times$720  \\
Pal3-6  (H4)  & Feb 11 1999 & 1$\times$600  \\		 
Pal3-8  (H5)  & Feb 12 1999 & 1$\times$600  \\
Pal3-15 (H6)  & Feb 11 1999 & 1$\times$960  \\
Pal3-16 (H8)  & Feb 12 1999 & 1$\times$1080 \\
Pal3-17 (H9)  & Feb 12 1999 & 1$\times$1200 \\
Pal3-18 (H10) & Feb 11 1999 & 1$\times$1200 \\
Pal3-21 (---) & Feb 11 1999 & 1$\times$1500 \\
Pal3-22 (H12) & Feb 12 1999 & 1$\times$1200 \\
Pal3-24 (H13) & Feb 12 1999 & 1$\times$1500 \\
Pal3-26 (H14) & Feb 12 1999 & 1$\times$1500 \\
Pal3-29 (H15) & Mar 10 1999 & 1$\times$1800 \\
Pal3-35 (H19) & Feb 12 1999 & 1$\times$1800 \\
Pal3-36 (H16) & Feb 12 1999 & 1$\times$1800 \\
Pal3-38 (H22) & Feb 12 1999 & 1$\times$1800 \\
Pal3-40 (H21) & Mar 10 1999 & 1$\times$2400 \\
Pal3-41 (H23) & Mar 10 1999 & 1$\times$2400 \\
\hline
\multicolumn{3}{c}{MIKE} \\
\hline                    
Pal3-2 (H2)   & Dec 31 2008, Jan 01 2009 & 4$\times$3600\\
Pal3-3 (H3)   & Jan 01 2008, Jan 02 2009 & 3$\times$3600\\
Pal3-5 (---)  & Jan 02 2008 & 3$\times$3600\\
Pal3-6 (H4)   & Jan 03 2008 & 4$\times$3600\\
\hline                                    
\end{tabular}
\\$^a$IDs preceded by ``H'' are taken from Hilker  (2006).
\end{table}

The data were reduced using the MAKEE\footnote{MAKEE was developed by T. A. Barlow specifically for reduction of Keck HIRES data. 
It is freely available on the World Wide Web at the Keck Observatory home page, \tt{http://www2.keck.hawaii.edu/inst/hires/makeewww}} 
data reduction package. 
Since our spectra were obtained with the original purpose of studying the internal cluster dynamics (C\^ot\'e et al. 2002), 
the exposure times -- which were chosen adaptively on the basis of target magnitude -- have low signal-to-noise (S/N) ratios. 
Thus, the spectra are adequate for the measurement of accurate radial velocities but not for abundance analyses of individual stars.
As a result of our observation and reduction strategy, we reach  S/N ratios of 4--7 per pixel in the order containing H$\alpha$. 
For the present study, we  stack the individual spectra to enhance the S/N ratio (Sect.~3.3)  and to perform an effective, integrated abundance analysis 
(see also McWilliam \& Bernstein 2008).  

Radial velocities of the individual targets were measured from a cross correlation against a synthetic spectrum of a red giant with 
stellar parameters representative of the Pal 3 target stars (see also Sect.~3.1), covering the whole wavelength range, but excluding the spectral gaps. 
We excluded stars deviating by more than 2$\sigma$ from the cluster's mean radial velocity of $\langle v_r \rangle \approx $ 92 km\,s$^{-1}$. 
A detailed account of the dynamics of Pal 3 will be given in a separate paper. 
%
%
\subsection{MIKE spectra}
Given the large distance of Pal~3 at (m$-$M)$_{\rm V,0} = 19.92$ mag (91.2 kpc; cf. Stetson et al. 1999; Hilker 2006), 
the stars on the upper RGB are faint at V$\sim$18.2--18.7 mag (Fig.~1).  
However,  we have shown in Koch et al. (2008a) that it is possible to target faint RGB stars in distant systems down 
to V$\sim$19 in reasonable integration times, with sufficient S/N ratios to investigate chemical abundances. 
Thus, we chose to observe the four brightest red giants using the Magellan Inamori Kyocera Echelle (MIKE) 
spectrograph at the 6.5-m Magellan2/Clay Telescope for detailed abundance analyses.
These targets were selected from the radial velocity member list based on our Keck run (Sect.~2.1); 
three of these stars are also included in Hilker (2006), who lists photometrically selected RGB and AGB member 
stars (Table~1). 

Our data were collected over four nights in January 2009. By using a slit width of 0.7$\arcsec$ and binning the 
CCD pixels by 2$\times$2, we obtained a resolving power of R$\sim$34000. 
Our data come from the red and blue sides of the instrument, which cover the wavelength range of 3340--9150\AA, 
although we will primarily use the red wavelength region above $\sim$4900\AA\ in our analysis. 
Each star was typically exposed for 3--4 hours, which we split into 1-hour exposures to facilitate cosmic ray removal. 
On average, the seeing was 1$\arcsec$ with individual exposures as high as 1.5$\arcsec$. 

The data were processed within the pipeline reduction package of Kelson (2000; 2003), which comprises  
flat field division, order tracing from quartz lamp flats, and 
wavelength calibration using built-in Th-Ar lamp exposures that were taken immediately following each science exposure. 
Continuum-normalisation was performed by dividing the extracted spectra by a high-order polynomial fitted to a spectrum of 
the essentially line-free hot rotating star HR~9098. 
Our MIKE spectra have S/N ratios of 30--40 per pixel as measured from the peak of the order containing H$\alpha$. 
\begin{table*}[htb]
\caption{Properties of the targeted member stars.}             
\centering          
\begin{tabular}{ccccccc}     
\hline\hline       
& $\alpha$  & $\delta$ & V$_0$ & (B$-$V)$_0$ & (V$-$K)$_0$ & S/N$^a$  \\
\raisebox{1.5ex}[-1.5ex]{ID} & (J2000.0) & (J2000.0)  & [mag] & [mag] & [mag] & [pixel$^{-1}$] \\
\hline
Pal3-2    & 10 05 31.57 & +00 04 17.0 & 18.11 & 1.13 &  2.73 & 7 / 44 \\
Pal3-3    & 10 05 37.08 & +00 04 27.9 & 18.22 & 1.11 &  2.57 & 7 / 32  \\
Pal3-5    & 10 05 30.70 & +00 04 04.2 & 18.35 & 1.07 &  2.52 & 7 / 36 \\
Pal3-6$^b$& 10 05 29.71 & +00 05 39.2 & 18.50 & 0.95 &  3.47 & 7 / 31 \\
Pal3-8    & 10 05 31.05 & +00 04 17.3 & 18.64 & 1.01 &  3.45 &  6 \\
Pal3-15   & 10 05 34.63 & +00 04 06.9 & 19.20 & 0.89 & \dots &  5 \\
Pal3-16   & 10 05 31.84 & +00 04 16.5 & 19.24 & 0.89 & \dots &  6 \\
Pal3-17   & 10 05 32.87 & +00 04 36.1 & 19.32 & 0.85 & \dots &  6 \\
Pal3-18   & 10 05 33.19 & +00 03 57.1 & 19.30 & 0.90 & \dots &  6 \\
Pal3-21   & 10 05 34.04 & +00 05 13.5 & 19.43 & 0.88 & \dots &  7 \\
Pal3-22   & 10 05 31.12 & +00 04 16.9 & 19.49 & 0.89 & \dots &  6 \\
Pal3-24   & 10 05 32.91 & +00 04 26.1 & 19.69 & 0.86 & \dots &  5 \\
Pal3-26   & 10 05 32.59 & +00 04 32.3 & 19.77 & 0.83 & \dots &  6 \\
Pal3-29   & 10 05 32.73 & +00 04 36.7 & 19.85 & 0.84 & \dots &  5 \\
Pal3-35   & 10 05 30.25 & +00 03 37.6 & 20.02 & 0.81 & \dots &  6 \\
Pal3-36   & 10 05 29.92 & +00 04 53.9 & 20.03 & 0.80 & \dots & 5 \\
Pal3-38   & 10 05 28.79 & +00 03 45.5 & 20.10 & 0.80 & \dots &  4 \\
Pal3-40   & 10 05 28.83 & +00 04 36.5 & 20.11 & 0.79 & \dots &  5 \\
Pal3-41   & 10 05 29.97 & +00 04 03.8 & 20.14 & 0.79 & \dots &  5 \\
\hline                  
\end{tabular}
\\$^a$The second number for the first 4 stars gives the S/N from the MIKE spectra.
\\$^b$AGB star (Hilker 2006).
\end{table*}
\section{Abundance analysis}
We begin with an analysis of our high-S/N MIKE spectra. We shall return to the integrated analysis of the co-added HIRES spectra 
in the next Section.
\subsection{Line list}
We derive chemical element abundances through a standard equivalent (EW) analysis that closely follows the procedures outlined in 
Koch \& McWilliam (2008) and Koch et al. (2008a, 2008b). 
We used the 2002 version of the stellar abundance code MOOG (Sneden 1973).  
The line list for this work is identical to the one we used in Koch et al. (2008a), which, in turn, was assembled from 
various sources (see Koch et al. 2008a,b; Koch \& McWilliam 2008; and references therein). Transitions for some heavy elements (Zr, La, Ce, Dy) were 
supplemented with data from Shetrone et al. (2003); Sadakane et al. (2004) and Yong et al. (2005).  Due to the current poorly-determined absolute 
abundance scale of the heavy elements in Arcturus (particularly the neutron capture elements), we opted to use the laboratory 
$gf$ values for our lines available in the literature (Koch et al. 2008a,b and references therein) rather than carrying out a differential abundance
analysis relative to this reference star (cf. Koch \& McWilliam 2008). The EWs were measured by fitting a Gaussian profile to the absorption lines 
using IRAF's {\em splot}. 
The final line lists are given in Table~3 for EWs from the MIKE spectra and in Table~4 for EWs from the co-added HIRES spectra.   
We comment on individual elements and transitions in Sect.~5.  
\begin{table*}[hbt]
\caption{Linelist for the MIKE spectra}             
\centering          
\begin{tabular}{cccccccc}     
\hline\hline       
& $\lambda$ & E.P. &  & EW (Pal3-2) & EW (Pal3-3) & EW (Pal3-5) & EW (Pal3-6)\\
\raisebox{1.5ex}[-1.5ex]{Element} & [\AA] & [eV]  &\raisebox{1.5ex}[-1.5ex]{log\,$gf$}  & [m\AA] & [m\AA] & [m\AA] & [m\AA]\\
\hline                    
[\ion{O}{I}] & 6300.31 & 0.00 & $-$9.819 & syn & \dots & syn &  syn \\
\ion{Na}{I} & 8183.26 & 2.10 &   0.230 & 120 & 136 & 114 &  88 \\
Na I  &  8183.26 & 2.10 &  0.230  &   120   &  136   &  114    & 88      \\ 
Na I  &  8194.79 & 2.10 & -0.470  &   143   &  160   &  134    & 124     \\ 
Mg I  &  5172.70 & 2.71 & -0.390  &   415   &  433   &  431    & 409     \\ 
Mg I  &  5528.42 & 4.35 & -0.357  &   160   &  177   &  167    & 167     \\ 
Mg I  &  5711.09 & 4.33 & -1.728  &   88    &  84    &  86     & 70      \\ 
\hline                  
\end{tabular}
\\Table~3 is available in its entirety in electronic form via the CDS.
\end{table*}
\begin{table*}[hbt]
\caption{Linelist for the co-added HIRES spectra}             
\centering          
\begin{tabular}{ccccccc}     
\hline\hline       
& $\lambda$ & E.P. &  & EW (All) & EW (Pal3-2,3,4,6) & EW (without Pal-2,3,4,6) \\
\raisebox{1.5ex}[-1.5ex]{Element} & [\AA] & [eV]  &\raisebox{1.5ex}[-1.5ex]{log\,$gf$}  & [m\AA] & [m\AA] & [m\AA] \\
\hline                    
Mg I & 5172.70 &  2.71 &  -0.390 & \dots &   475 & \dots \\
Mg I & 5528.42 &  4.35 &  -0.357 & \dots &   176 & \dots \\
Mg I & 5711.09 &  4.33 &  -1.728 &    83 &    85 & \dots \\
Al I & 6696.03 &  3.14 &  -1.347 &    36 &    40 & \dots \\
Si I & 5684.48 &  4.95 &  -1.650 & \dots &    34 & \dots \\
Si I & 5708.41 &  4.95 &  -1.470 &    63 &    77 &    74 \\
Si I & 5948.55 &  5.08 &  -1.230 &    41 &    72 &    42 \\
Si I & 6142.48 &  5.62 &  -0.920 & \dots &    22 & \dots \\
Si I & 6155.13 &  5.61 &  -0.750 &    28 & \dots &    30 \\
\hline                  
\end{tabular}
\\Table~4 is available in its entirety in electronic form via the CDS.
\end{table*}

We accounted for the effects of hyperfine structure for the stronger lines of the odd-Z elements Mn I, Cu I, Ba II, La II, and Eu II, 
using data for the splitting from McWilliam et al. (1995). However, the hyperfine splitting for Sc II, V I, Co I, and Y II was negligible 
for the weak lines employed in our study and we ignored this effect for these elements. 
Finally, we placed our abundances on the Solar scale of 
Asplund et al. (2005), except for iron, for which we adopted $\log\,\varepsilon_{\odot}$(Fe)=7.50 as an average of the values found in the 
literature during the past  years (see also McWilliam \& Bernstein 2008).  
\subsection{Stellar atmospheres}
Throughout our analysis we interpolated the model atmospheres from the updated grid of the Kurucz\footnote{\tt http://cfaku5.cfa.harvard.edu/grids.html} 
one-dimensional 72-layer, plane-parallel, line-blanketed models 
without convective overshoot and assuming local thermodynamic equilibrium (LTE) 
for all species. Our models incorporated the new $\alpha$-enhanced opacity distribution functions, 
AODFNEW (Castelli \& Kurucz 2003)\footnote{See {\tt http://wwwuser.oat.ts.astro.it/castelli}.}. 
This seems a reasonable choice since the majority of the metal poor Galactic halo GCs and field stars are enhanced in the $\alpha$-elements 
by $\approx +0.4$ dex, and we would expect Pal~3 to follow this trend (see also Fig.~1; Sects.~4, 5.4).   

Photometric temperatures were obtained from the stellar (V$-$K) colours using the temperature-colour calibrations of Ram\'irez \& Mel\'endez (2005)  
with K-band photometry from 2MASS (Cutri et al. 2003). A reddening of E(B$-$V)=0.04 (Stetson et al. 1999; Hilker 2006),  the  
extinction law of Winkler (1997),  and an estimated mean metallicity of $-1.6$ dex from previous photometric studies 
were also adopted.  
Given the large uncertainties in the 2MASS K-magnitudes of our faint targets, the resulting T$_{\rm eff}$ (V$-$K)  estimates
have formal uncertainties of  $\pm$170 K on average. 
In addition, we derived T$_{\rm eff}$ from our LRIS   
$BV$ photometry by employing the temperature calibrations of Alonso et al. (1999). 
In practice, we adopted the (V$-$K)-based values as initial temperatures for our stellar atmospheres 
for all stars, with one exception: the AGB star Pal3-6 exhibits too red a (V$-$K) colour  
(possibly due to an erroneous K-band magnitude or an unresolved blend in the 2MASS)   
that  lead to a T$_{\rm eff}$ lower by 600 K compared to the (B$-$V)-based value. 
We then derived spectroscopic temperatures by demanding excitation equilibrium; that is, by requiring there be no trend in the abundance from 
the Fe~I lines with excitation potential. 
This procedure typically yields T$_{\rm eff}$ accurate to within $\pm$100 K, based on the range of reasonable slopes. 
As a result, the spectroscopic temperatures are slightly higher than the photometric values, by 80$\pm$17 K on average. 
In practice, we adopted the spectroscopic T$_{\rm eff}$ to enter our atmospheres, as this provides the most reliable and independent determination. 
In Table~5 we list the atmospheric parameters of the red giants analysed with MIKE.
\begin{table}[htb]
\caption{Atmospheric parameters of the MIKE stars}             
\centering          
\begin{tabular}{cccccc}     
\hline\hline       
& \multicolumn{3}{c}{T$_{\rm eff}$ [K]} &  & $\xi$  \\
\cline{2-4}
\raisebox{1.5ex}[-1.5ex]{ID} &(B$-$V)& (V$-$K) & (spec)  & \raisebox{1.5ex}[-1.5ex]{log\,$g$} & [km\,s$^{-1}$]  \\
\hline                    
Pal3-2 & 4360 & 4370 & 4470 & 1.08 & 2.10 \\
Pal3-3 & 4390 & 4490 & 4520 & 1.16 & 2.30 \\ 
Pal3-5 & 4440 & 4530 & 4630 & 1.28 & 2.20  \\
Pal3-6 & 4610 & 3960 & 4700 & 1.38 & 2.20  \\
\hline                  
\end{tabular}
\end{table}

Photometric gravities were derived from the basic stellar structure equations (e.g., Koch \& McWilliam 2008)  
using the above temperatures, our V-band photometry, and adopting a dereddened distance modulus  for Pal 3 of 19.92 mag, which 
was found to yield a satisfactory fit to our CMD in Fig.~1  
(cf. Stetson et al. 1999; Hilker 2006).  Comparison of the colours and magnitudes of the Pal 3 red giants 
against a set of Teramo isochrones (Pietrinferni et al. 2004) with an age of 10 Gyr and an  [Fe/H] of $-$1.6 dex (Stetson et al. 1999; Hilker 2006)
indicates an average  stellar mass of 0.84 M$_{\odot}$ for the red giants and 0.78 M$_{\odot}$ for the AGB star, which we  
adopted for the log\,$g$ determinations.  
Uncertainties in the distance and photometry imply an average error in our gravities of 0.12 dex.
We note that ionisation equilibrium is not fulfilled in our stars, where we find  a mean deviation of the neutral 
and ionised species of [Fe\,{\sc i}/Fe\,{\sc ii}]=$-0.07\pm$0.01 dex, while the [Ti\,I/II] ratio is 0.02$\pm0.04$ dex. 

We did not attempt to enforce equilibrium by marginal changes in log $g$ (see also Koch \& McWilliam 2008), but note that an increase of log\,$g$ by 
0.18$\pm$0.03 dex would settle the discrepancy at an [Fe/H] higher by 0.05$\pm$0.01dex on average (see also Sect.4; Table~6). 
A mild increase of the temperature scale by  37$\pm$3 K would also reinstall the ionisation balance at marginally higher metallicities. 

We then determined microturbulent velocities, $\xi$, from the plot of abundances versus EW of the iron lines, 
which yields $\xi$, typically to within 0.2 km\,s$^{-1}$. 
The resulting  microturbulent velocity values of $\sim$2.2 km\,s$^{-1}$ are somewhat higher 
than those found in red giants with similar parameters (e.g., Cayrel et al. 2004; Yong et al. 2005). 
This is likely an artifact of the relatively low S/N ratios of our spectra and the asymmetrical abundance errors 
of the stronger lines (Magain 1984). However, our analysis reassures us that the high microturbulent 
velocities from the EW plot are appropriate for our spectra. 

Since there is no prior knowledge of the individual stellar metallicities, we initially 
adopted the cluster mean of $-1.6$ dex (Hilker 2006, and references therein) as input for the model atmospheres. 
An independent estimate can be reached from two indicators. 

First, our spectra contain the near-infrared calcium triplet (CaT) lines at 8498, 8452, and 8662\AA;  
these lines are a well-calibrated indicator of the metallicities of red giants in Galactic GCs (Armandroff \& Zinn 1988; 
Rutledge et al. 1997a,b; Carretta \& Gratton 1997). 
The measurement of the CaT from high-resolution data, however, should be treated with caution since the wings are 
strong, and any blends with weaker sky residuals, telluric lines, or neutral metal lines will affect their 
shapes; simplistic line profiles usually fail to fit reliably the CaT lines in very high-resolution spectra.  
Furthermore, these lines form in the upper chromospheres and  are difficult to model reliably (e.g., McWilliam et al. 1995; 
Battaglia et al. 2008) and  setting their continua is not straightforward in our type of spectra. 
Nonetheless, for a simple order-of-magnitude estimate, we fit the line profiles of the CaT lines 
as any other absorption line with a Gaussian using {\em splot}. 
The measured EWs were then converted to metallicities on the scale of Carretta \& Gratton (1997) using the 
calibrations of Rutledge et al. (1997a,b): 
\begin{equation}
{\rm [Fe/H]}_{\rm CaT} = -2.66+0.42\,\left[\Sigma W+0.64\,(V-V_{\rm HB})\right],  
\end{equation}
where $\Sigma W=0.5W_{8498}+W_{8452}+0.6W_{8662}$ is the weighted sum of the EWs, and $V_{\rm HB}$ 
denotes the horizontal branch magnitude, which we adopted as 20.5 mag (Hilker 2006). 
As a result, we find a mean CaT metallicity  [Fe/H]$_{\rm CaT}$ of our stars of $-1.60\pm0.02$ dex,  
which is in good agreement with both photometric estimates ([Fe/H]$=-1.57$ dex;  Stetson et al. 1999) 
and the results of low-resolution spectroscopy ([Fe/H]$_{\rm CaT}=-1.57\pm0.19$ dex; Armandroff et al. 1992), as well as  in 
fair agreement with the value of $-$1.70$\pm$0.20 dex reported by Ortolani \& Gratton (1989).

As a second metallicity indicator, we integrated the Mg I lines at 5167, 5173\AA\ to calibrate [Fe/H] on the  scale of Carretta \& Gratton (1997) as
\begin{equation}
{\rm [Fe/H]}_{\rm Mg I} =  -2.11 +1.76\,\left[ \Sigma {\rm Mg}   + 0.079\,(V - V_{\rm HB})\right]. 
\end{equation}
Here, $\Sigma {\rm Mg}=0.547\,(W_{5167}+W_{5173})$ denotes the Mg index as defined and calibrated in Walker et al. (2007). 
The respective errors were obtained from the formalism of Cardiel et al. (1998). 
We find a mean [Fe/H]$_{\rm Mg~I}$ of our stars of $-1.57\pm0.04$ dex, which is in good agreement with the CaT value. 
Both the CaT and Mg~I values are listed in Table~7, but we 
emphasise again that these values are meant as initial estimates of the cluster metallicity rather than reliable measures of its abundance scale.

To conclude, 
we used the abundance of the Fe~I lines as input metallicity for the next iteration as we iterated the parameter derivation simultaneously in all parameters until convergence was reached. 
\subsection{Co-added HIRES spectra}
Since the HIRES data were originally taken with the sole intention of studying the cluster dynamics (C\^ot\'e et al. 2002), the S/N ratio 
of the spectra is low ($\le$7) and does not allow for 
abundance measurements of the individual stars. Instead, we coadded the spectra to enhance their S/N in order  
to yield integrated element ratios (cf. McWilliam \& Bernstein 2008).  As the high-resolution, high-S/N 
MIKE spectra already resulted in a wealth of abundance information, we followed three procedures so as not to bias the spectral co-addition 
against these known abundances. First, we co-added all spectra, {\em excluding} those stars targeted with MIKE. Second, 
spectra of all 19 radial velocity members were stacked, {\em including} the four MIKE stars. 
Finally, {\em only} the four stars analysed in the previous section were combined 
into a higher-S/N spectrum. The latter allows us to compare the results of the individual spectra versus the co-added ones and to look for 
potential systematic differences between the two methods. Using these three co-adddition schemes, we emulated total S/N ratios of 22, 26,  
and 14 in the H$\alpha$-order. 

In practice, the spectra were Doppler-shifted and average-combined after weighting by their individual S/N ratios. In order to yield abundances, we measured the 
EWs from the stacked spectra using the same methods and line lists as in Sect. 3.1 (note, however, the reduced spectral range compared to MIKE). 
A few of the lines that were measured in the individual MIKE spectra had to be discarded from the co-added line list 
because they were too weak to measure at the comparatively low S/N. Strong lines (such as a few Mg or Cr lines)  for which the coaddition 
rendered  asymmetric profiles and/or line wings too strong to be reliably measured were also excluded. 
The goal is then to compare these to theoretical  EWs from synthetic spectra. To generate these, we computed individual spectra using 
model atmospheres that represent each star's stellar parameters.
 
The parameters for our stars were obtained following the procedures described above. 
Since the targets are relatively faint, they mostly fall below the magnitude limit of the 2MASS (Table~2). Thus, we must rely exclusively on the LRIS 
$BV$ photometry to obtain T$_{\rm eff}$. 
Surface gravities were derived, as before, from the stars' photometry.  
Since the parameters of the MIKE stars follow the trend outlined by the metal poor halo stars 
from Cayrel et al. (2004), we derived microturbulent velocities from a linear fit to their data as $\xi = 6.55 - 9.62\,\times10^{-4}\,$T$_{\rm eff}$.  
Histograms  
of the  parameters for our HIRES sample are presented in Fig.~2. 
\begin{figure}[htb]
\centering
\includegraphics[width=1\hsize]{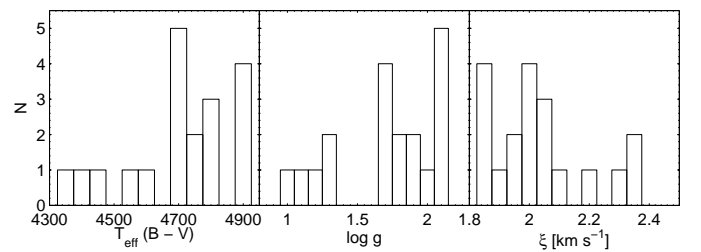}
\caption{Distribution of photometric (B$-$V) temperatures, gravities and microturbulence of the HIRES targets.}
\end{figure}

The spectral range of our HIRES setup does not contain the near-infrared CaT, but we can obtain metallicity estimates from 
the Mg I index (see Eq.~2). Owing to the low S/N ratios, the uncertainties on the resulting metallicity are inevitably 
large,  typically up to 0.5 dex, but we retain this method for an initial, order-of-magnitude estimate.  
Doing so, we find a mean [Fe/H]$_{\rm Mg I}$ of $-1.60$ dex on 
the scale of Carretta \& Gratton (1997) with a 1$\sigma$ scatter of 0.28 dex. 

With the resulting atmospheres in hand, we computed theoretical EWs for the transitions in our line list using MOOG's {\em ewfind} driver 
and combined them into a mean value, $<$$EW$$>$, using the same weighting scheme as for the observations: 
\begin{equation} 
<\!EW\!> \, = \, \frac{ \sum_{i=1}^N\,w_i\,EW_i\, }{ \,\sum_{i=1}^N\,w_i}, 
\end{equation}
where the weights $w_i$ are proportional to the S/N ratios as in the case of co-adding the observed spectra.  
A comparison of $<$$EW$$>$ to the observed co-added EW then yields the cluster's integrated abundance ratio for each element. 
Note that this method presupposes that there is no significant abundance scatter present along the RGB and all stars 
have the same mean abundances for all chemical elements. 
\section{Abundance errors}
The systematic errors on our abundances from MIKE were determined by computing eight new stellar atmospheres, in which 
each stellar parameter (T$_{\rm eff}$, log\,$g$, $\xi$, [$M$/H]) was varied by its typical uncertainty (as estimated in Sect.~3.1).  
In addition, we re-ran the analysis using the solar-scaled opacity distributions, ODFNEW, which corresponds to an uncertainty 
in the input [$\alpha$/Fe] ratio of 0.4 dex. With these new atmospheres, new element ratios were determined and we list in 
Table~6  the abundance differences from those derived using  the best-fit atmospheric parameters. 
This procedure was performed for the stars Pal3-2 and Pal3-6, which cover the full range in T$_{\rm eff}$. 
\begin{table*}
\caption{Error analysis for the red giants Pal3-2 (RGB) and Pal3-6 (AGB).}
\centering          
\begin{tabular}{rrrrrrrrrr}
\hline
\hline
& \multicolumn{2}{c}{$\Delta$T$_{\rm eff}$} & \multicolumn{2}{c}{$\Delta\,\log\,g$} & \multicolumn{2}{c}{$\Delta\xi$} 
& \multicolumn{2}{c}{$\Delta$[M/H]} &   \\
\raisebox{1.5ex}[-1.5ex]{Ion}  & $-$100\,K  & +100\,K & $-$0.2\,dex & +0.2\,dex & $-$0.2\,km\,s$^{-1}$ & +0.2\,km\,s$^{-1}$ & $-$0.1\,dex & +0.1\,dex &  \raisebox{1.5ex}[-1.5ex]{ODF} \\
\hline
\ion{Fe}{I}  & $-$0.13 &    0.14 &    0.01 & $<$0.01 & 0.10    & $-$0.10 &    0.01 & $-$0.01 &    0.01 \\ 
\ion{Fe}{II} &    0.10 & $-$0.08 & $-$0.10 &	0.09 & 0.06    & $-$0.06 & $-$0.03 &	0.02 & $-$0.09 \\ 
\ion{Na}{I}  & $-$0.08 &    0.06 & $<$0.01 &	0.01 & 0.06    & $-$0.06 &    0.01 & $-$0.01 &    0.01 \\ 
\ion{Mg}{I}  & $-$0.09 &    0.07 &    0.04 & $-$0.04 & 0.07    & $-$0.07 &    0.01 & $<$0.01 &    0.02 \\ 
\ion{Si}{I}  & $<$0.01 & $<$0.01 & $-$0.02 &	0.02 & 0.02    & $-$0.02 & $<$0.01 &	0.01 & $-$0.01 \\ 
\ion{K}{I}   & $-$0.16 &    0.14 & $<$0.01 & $<$0.01 & 0.12    & $-$0.12 &    0.02 & $-$0.02 &    0.02 \\ 
\ion{Ca}{I}  & $-$0.11 &    0.09 &    0.02 & $-$0.02 & 0.08    & $-$0.08 &    0.01 & $-$0.01 &    0.03 \\ 
\ion{Sc}{II} & $-$0.01 & $<$0.01 & $-$0.08 &	0.08 & 0.02    & $-$0.04 & $-$0.03 &	0.03 & $-$0.08 \\ 
\ion{Ti}{I}  & $-$0.22 &    0.21 & $<$0.01 & $-$0.02 & 0.09    & $-$0.09 &    0.02 & $-$0.02 &    0.02 \\ 
\ion{Ti}{II} & $<$0.01 & $-$0.02 & $-$0.06 &	0.09 & 0.10    & $-$0.08 & $-$0.02 &	0.02 & $-$0.07 \\ 
\ion{V}{I}   & $-$0.18 &    0.18 &    0.02 & $-$0.02 & 0.02    & $-$0.01 &    0.01 & $<$0.01 &    0.04 \\ 
\ion{Cr}{I}  & $-$0.17 &    0.17 &    0.02 & $-$0.02 & 0.10    & $-$0.09 &    0.02 & $-$0.01 &    0.04 \\ 
\ion{Mn}{I}  & $-$0.10 &    0.10 &    0.02 & $-$0.02 & 0.01    & $-$0.01 &    0.01 &	0 00 &    0.03 \\ 
\ion{Co}{I}  & $-$0.13 &    0.14 & $<$0.01 & $<$0.01 & 0.02    & $-$0.01 & $<$0.01 & $<$0.01 &    0.02 \\ 
\ion{Ni}{I}  & $-$0.10 &    0.10 & $-$0.01 &	0.01 & 0.04    & $-$0.04 & $<$0.01 & $<$0.01 &    0.01 \\ 
\ion{Cu}{I}  & $-$0.14 &    0.12 & $<$0.01 & $-$0.02 & 0.06    & $-$0.06 &    0.01 & $-$0.01 &    0.03 \\ 
\ion{Sr}{II} & $-$0.04 &    0.02 & $-$0.04 &	0.02 & 0.02    & $-$0.04 & $-$0.03 &	0.03 & $-$0.09 \\ 
\ion{Y}{II}  & $-$0.02 &    0.02 & $-$0.07 &	0.08 & 0.05    & $-$0.04 & $-$0.03 &	0.03 & $-$0.08 \\ 
\ion{Zr}{II} & $-$0.02 & $<$0.01 & $-$0.08 &	0.08 & $<$0.01 & $-$0.02 & $-$0.03 &    0.02 & $-$0.07 \\ 
\ion{Ba}{II} & $-$0.05 &    0.04 & $-$0.06 &	0.10 & 0.17    & $-$0.15 & $-$0.02 &	0.03 & $-$0.09 \\ 
\ion{La}{II} & $-$0.05 &    0.03 & $-$0.07 &	0.08 & 0.01    & $-$0.02 & $-$0.03 &	0.03 & $-$0.09 \\ 
\ion{Nd}{II} & $-$0.04 &    0.03 & $-$0.08 &	0.08 & 0.03    & $-$0.04 & $-$0.03 &	0.03 & $-$0.08 \\ 
\ion{Eu}{II} & $-$0.01 & $<$0.01 & $-$0.08 &	0.09 & 0.01    & $-$0.01 & $-$0.03 &	0.03 & $-$0.09 \\ 
\hline
\ion{Fe}{I}  & $-$0.14 &    0.13 &    0.01 & $-$0.01 &    0.09 & $-$0.08 &    0.01 & $-$0.01 &    0.04 \\ 
\ion{Fe}{II} &    0.02 & $-$0.03 & $-$0.07 &	0.06 &    0.06 & $-$0.06 & $-$0.02 &	0.02 & $-$0.05 \\ 
\ion{Na}{I}  & $-$0.07 &    0.06 &    0.01 & $-$0.01 &    0.01 & $-$0.06 &    0.01 & $-$0.01 &    0.03 \\ 
\ion{Mg}{I}  & $-$0.08 &    0.08 &    0.05 & $-$0.04 &    0.06 & $-$0.05 &    0.01 & $<$0.01 &    0.02 \\ 
\ion{Si}{I}  & $-$0.04 &    0.02 & $<$0.01 & $<$0.01 &    0.02 & $-$0.02 & $<$0.01 & $<$0.01 &    0.01 \\ 
\ion{K}{I}   & $-$0.12 &    0.10 & $<$0.01 & $-$0.02 &    0.08 & $-$0.06 &    0.02 & $-$0.01 &    0.04 \\ 
\ion{Ca}{I}  & $-$0.09 &    0.08 &    0.02 & $-$0.01 &    0.05 & $-$0.05 &    0.01 & $<$0.01 &    0.03 \\ 
\ion{Sc}{II} & $-$0.01 & $<$0.01 & $-$0.08 &	0.08 &    0.04 & $-$0.03 & $-$0.02 &	0.03 & $-$0.07 \\ 
\ion{Ti}{I}  & $-$0.18 &    0.17 &    0.02 & $-$0.01 &    0.06 & $-$0.05 &    0.02 & $-$0.01 &    0.05 \\ 
\ion{Ti}{II} & $-$0.01 & $-$0.02 & $-$0.07 &	0.06 &    0.08 & $-$0.08 & $-$0.02 &	0.02 & $-$0.07 \\ 
\ion{V}{I}   & $-$0.18 &    0.16 &    0.02 & $-$0.02 & $<$0.01 & $-$0.02 &    0.01 & $-$0.01 &    0.04 \\ 
\ion{Cr}{I}  & $-$0.14 &    0.14 &    0.02 & $-$0.02 &    0.07 & $-$0.06 &    0.02 & $-$0.01 &    0.05 \\ 
\ion{Mn}{I}  & $-$0.11 &    0.10 &    0.02 & $<$0.01 &    0.01 & $<$0.01 &    0.01 & $<$0.01 &    0.03 \\ 
\ion{Co}{I}  & $-$0.15 &    0.15 &    0.01 & $<$0.01 &    0.02 & $-$0.01 &    0.01 & $<$0.01 &    0.03 \\ 
\ion{Ni}{I}  & $-$0.12 &    0.11 & $<$0.01 & $<$0.01 &    0.03 & $-$0.04 &    0.01 & $<$0.01 &    0.03 \\ 
\ion{Cu}{I}  & $-$0.14 &    0.14 & $<$0.01 & $-$0.02 &    0.02 & $-$0.02 &    0.01 &	0.01 &    0.03 \\ 
\ion{Sr}{II} & $-$0.04 &    0.04 & $-$0.02 &	0.04 &    0.04 & $-$0.04 & $-$0.02 &	0.03 & $-$0.07 \\ 
\ion{Y}{II}  & $-$0.02 &    0.02 & $-$0.08 &	0.06 &    0.06 & $-$0.04 & $-$0.02 & $-$0.02 & $-$0.06 \\ 
\ion{Ba}{II} & $-$0.04 &    0.03 & $-$0.07 &	0.08 &    0.16 & $-$0.13 & $-$0.02 &	0.02 & $-$0.07 \\ 
\ion{La}{II} & $-$0.04 &    0.04 & $-$0.06 &	0.08 &    0.08 & $-$0.02 & $-$0.02 & $-$0.02 & $-$0.06 \\ 
\ion{Nd}{II} & $-$0.04 &    0.04 & $-$0.06 &	0.08 &    0.08 & $-$0.02 & $-$0.02 & $-$0.02 & $-$0.06 \\ 
\ion{Eu}{II} & $-$0.01 &    0.02 & $-$0.07 &	0.08 &    0.01 & $<$0.01 & $-$0.03 &	0.03 & $-$0.07 \\ 
\hline                  
\end{tabular}
\end{table*}

As a measure for the total systematic uncertainty, we sum in quadrature the contributions from each parameter, although 
we note that this yields conservative upper limits; the real underlying errors will be smaller due to the covariances 
of the atmosphere parameters, in particular, that between temperature and gravity (e.g., McWilliam et al. 1995; Johnson 2002; 
Koch et al. 2008b).  
As Table~6 shows, the total error on the iron abundances is thus  0.13 dex for the neutral and 0.17 for the ionised   
species. The $\alpha$ elements are typically uncertain to within 0.15 dex, although there are differences between individual 
elements (i.e., smaller errors for Si and Ca, and errors on the [Ti/Fe] ratios that are slightly larger). Likewise, the iron peak and neutron 
capture elements show abundance errors of 0.13--0.17  dex, on average, with the exception of Ba, for which we find a slightly larger
uncertainty of $\sim$0.21 dex. 
As already noted by Ram\'irez \& Cohen (2003), the Ti I, V I, and Cr I abundance ratios show a strong dependence on T$_{\rm eff}$.  
This is explained by their low excitation potential -- at the temperatures of the stars studied here, these species begin to change from  fully ionised  to 
fully neutral. 
Ba II, on the other hand, shows a strong trend with microturbulence due to 
the generally strong lines. None of the measured abundance ratios is strongly affected by changes in the input metallicties [$M$/H], 
while changes in the [$\alpha$/Fe] ratio of the atmospheres lead to a larger impact on all ionised  species compared to the neutral stages. 

In Table~7 we list the 1$\sigma$ line-to-line scatter and number of lines that was used to derive the abundance ratios listed
in this table. This error component yields a measure of the random error accounting for the spectral noise, uncertainties in 
the atomic parameters, and insufficiencies in the atmosphere models themselves. 
For elements with many measurable lines, like Fe, Ca, Ti, or Ni, the systematic uncertainties will dominate, whereas for elements 
with only a few detectable transitions, the line-to-line scatter is the dominant error source. 
For those elements for which only one line was detectable, we
adopted an uncertainty of 0.10 dex. For all other elements, we assumed a minimum random error of 0.05 dex (e.g., Ram\'irez \& Cohen 2003). 
We will return to the question of intrinsic versus real stellar scatter in Sect.~6.1. 

In the case of the co-added HIRES analysis, systematic uncertainties are more difficult to evaluate as the errors on individual stellar parameters 
propagate through the weighted averaging of the spectra.  
However, as the 1$\sigma$ line-to-line scatter on those measurements in Table~8 indicates, the statistical error component is large and  
overwhelms any systematic dependence on the weighted and averaged stellar parameters. 
This can be due to the low spectral S/N, even after the co-addition. 
In particular, an accurate placement of the continuum is not easily achieved in the stacked low S/N spectra 
and can lead to random over- and underestimates of measured EWs. 
Accordingly, we adopted a minimum random abundance error of 0.10 dex for the HIRES results and assign an uncertainty of 0.15 dex if only 
one line could be measured. 
 \section{Abundance results}
Table~7 lists the abundance ratios relative to Fe I, except for all ionised  species and O I from the forbidden line, which we give relative to Fe II. 
These are also illustrated in the boxplots of Fig.~3 which show the median and interquartile ranges for each element's abundance ratio. 
\begin{table*}[htb]
\caption{Abundance results of the giants targeted with MIKE.}             
\centering          
\begin{tabular}{cccrcccrcccrcccr}     
\hline\hline       
& & Pal3-2 & & & & Pal3-3 & & & & Pal3-5 &  & & & Pal3-6 &\\
\cline{2-4}\cline{6-8}\cline{10-12}\cline{14-16}
\raisebox{1.5ex}[-1.5ex]{Element$^a$} & [X/Fe] & $\sigma$ & N &  & [X/Fe] & $\sigma$ & N &  & [X/Fe] & $\sigma$ & N &  & [X/Fe] & $\sigma$ & N \\
\hline                    
%
%
Fe$_{\rm CaT}^b$  & $-$1.56 & 0.14  & \dots & & $-$1.58 & 0.14 & \dots  & & $-$1.63 & 0.14 & \dots  & & $-$1.61 & 0.17 & \dots \\
Fe$_{\rm Mg I}^c$ & $-$1.51 & 0.13  & \dots & & $-$1.51 & 0.16 & \dots  & & $-$1.58 & 0.13 & \dots  & & $-$1.67 & 0.14 & \dots \\
Fe\,{\sc i}      & $-$1.59  &  0.24 & 141 & & $-$1.54 &  0.27 & 137 & & $-$1.56 &  0.27 &   139 & & $-$1.62 &  0.22 & 121 \\
Fe\,{\sc ii}     & $-$1.52  &  0.19 &	9 & & $-$1.50 &  0.16 &   6 & & $-$1.49 &  0.17 &    10 & & $-$1.52 &  0.21 &	8 \\
O\,{\sc i}   &    0.29  & \dots &	1 & &	\dots & \dots &\dots& &    0.30 & \dots &     1 & &    0.32 & \dots &   1  \\
Na\,{\sc i}      &    0.13  &  0.01 &	2 & &	 0.27 &  0.01 &   2 & &    0.13 &  0.02 &     2 & &    0.00 &  0.12 &	2 \\
Mg\,{\sc i}      &    0.36  &  0.11 &	3 & &	 0.37 &  0.08 &   3 & &    0.47 &  0.04 &     3 & &    0.48 &  0.08 &	3 \\
Si\,{\sc i}      &    0.44  &  0.08 &	5 & &	 0.46 &  0.10 &   4 & &    0.50 &  0.02 &     2 & &    0.57 &  0.12 &	2 \\
K\,{\sc i}       &    0.51  & \dots &	1 & &	 0.48 & \dots &   1 & &    0.58 & \dots &     1 & &    0.24 & \dots &	1 \\
Ca\,{\sc i}      &    0.31  &  0.12 &  14 & &	 0.27 &  0.08 &  13 & &    0.33 &  0.12 &    14 & &    0.30 &  0.14 &  12 \\
Sc\,{\sc ii}     &    0.26  &  0.08 &	5 & &	 0.40 &  0.07 &   4 & &    0.13 &  0.07 &     4 & &    0.21 &  0.04 &   6 \\
Ti\,{\sc i}      &    0.39  &  0.11 &  20 & &	 0.31 &  0.13 &   8 & &    0.37 &  0.12 &    12 & &    0.23 &  0.14  & 12 \\
Ti\,{\sc ii}     &    0.33  &  0.11 &	5 & &	 0.22 &  0.16 &   2 & &    0.19 &  0.10 &     4 & &    0.22 &  0.23  &  4 \\
V\,{\sc i}       &    0.13  &  0.07 &	8 & &	 0.10 &  0.11 &   7 & &    0.20 &  0.12 &     4 & &    0.10 & \dots &	1   \\
Cr\,{\sc i}      &    0.01  &  0.09 &  10 & &	 0.13 &  0.12 &   8 & &    0.15 &  0.15 &     6 & &    0.00 &  0.13 &	5   \\
Mn\,{\sc i}      & $-$0.32  &  0.07 &	3 & &	\dots & \dots &\dots& & $-$0.23 &  0.11 &     3 & & $-$0.37 &  0.14 &	3   \\
Co\,{\sc i}      &    0.11  &  0.05 &	5 & &	 0.22 &  0.11 &   4 & &    0.17 &  0.06 &     3 & &    0.15 &  0.06 &	2   \\
Ni\,{\sc i}      & $-$0.04  &  0.16 &  24 & &	 0.05 &  0.19 &  17 & & $-$0.03 &  0.14 &    15 & &    0.01 &  0.24 &  17   \\
Cu\,{\sc i}      & $-$0.26  & \dots &	1 & & $-$0.25 & \dots &   1 & & $-$0.22 & \dots &     1 & & $-$0.47 & \dots &	1   \\
Sr\,{\sc ii}     & $-$0.46  & 0.00 &	2 & & $-$0.42 & 0.08 &   2 & &   \dots & \dots & \dots & & $-$0.27 & 0.10  &   2   \\
Y\,{\sc ii}      & $-$0.03  &  0.04 &	2 & &	 0.12 &  0.04 &   2 & & $-$0.02 & \dots &     1 & & $-$0.08 & \dots &	1   \\
Zr\,{\sc ii}     &    0.03  & \dots &	1 & &	 0.32 & \dots &   1 & &    0.13 & \dots &     1 & &   \dots & \dots & \dots \\
Ba\,{\sc ii}     &    0.04  &  0.11 &	3 & & $-$0.02 &  0.19 &   3 & & $-$0.00 &  0.08 &     3 & &    0.02 &  0.05 &	3   \\
La\,{\sc ii}     &    0.40  &  0.05 &	3 & &	 0.36 & \dots &   1 & &    0.36 & \dots &     1 & &    0.44 & \dots &	1   \\
Nd\,{\sc ii}     &    0.31  &  0.02 &	2 & &	 0.31 & \dots &   1 & &    0.31 &  0.08 &     2 & &    0.29 & \dots &	1   \\
Eu\,{\sc ii}     &    0.74  &  0.01 &	2 & &	 0.61 &  0.11 &   2 & &    0.67 &  0.07 &     2 & &    0.80 &  0.18 &	2   \\
\hline                  
\end{tabular}
\begin{flushleft}
$^a$Ionised  species and O are given relative to Fe\,{\sc ii}.
\\$^b$Metallicity estimate based on the calcium triplet calibration of Rutledge 
et al. (1997a,b), on the metallicity scale of Carretta \& Gratton (1997).
\\$^c$Metallicity estimate based on the Mg I calibration of Walker 
et al. (2007), on the metallicity scale of Carretta \& Gratton (1997).
\end{flushleft}
\end{table*}
\begin{figure}[htb]
\centering
\includegraphics[width=1\hsize]{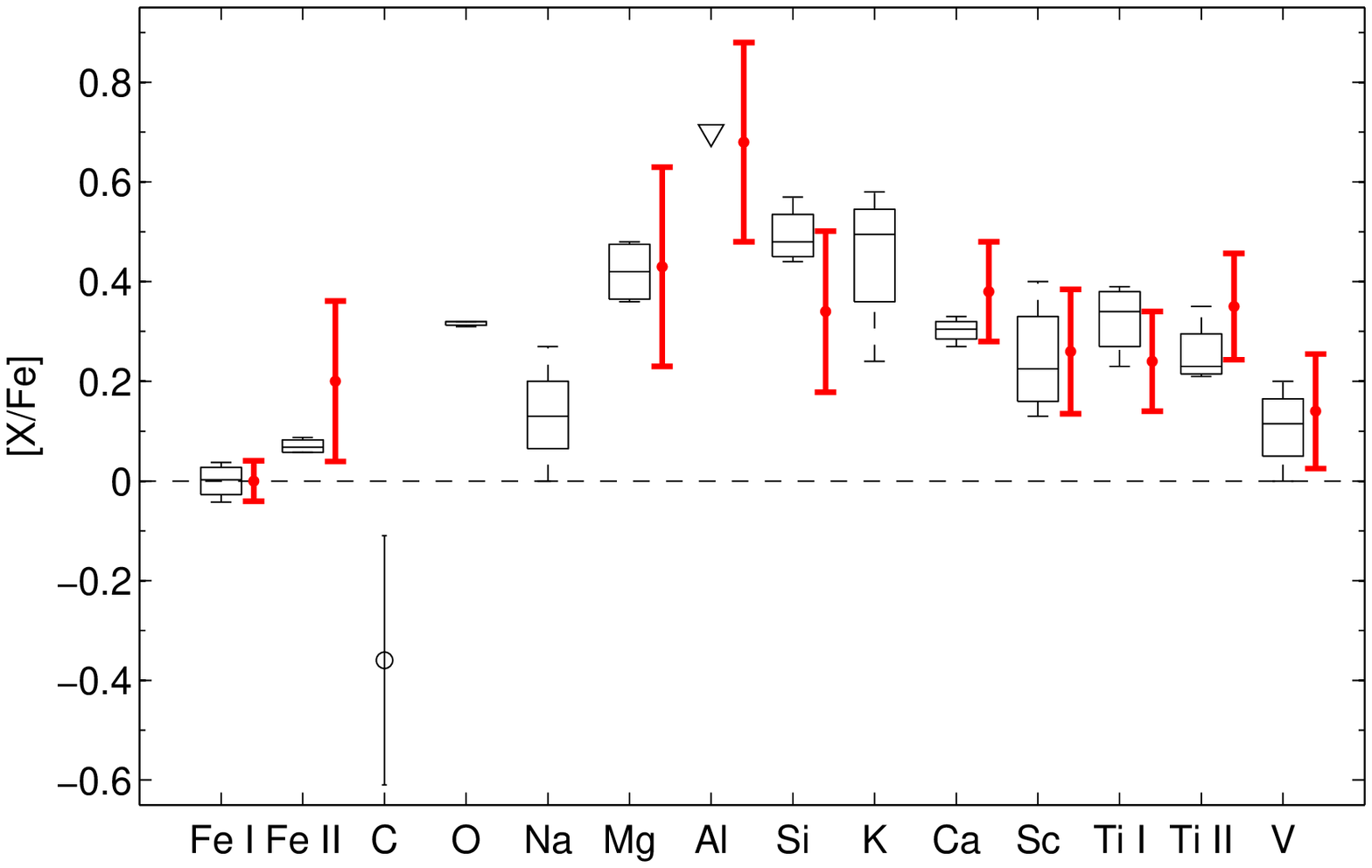}
\includegraphics[width=1\hsize]{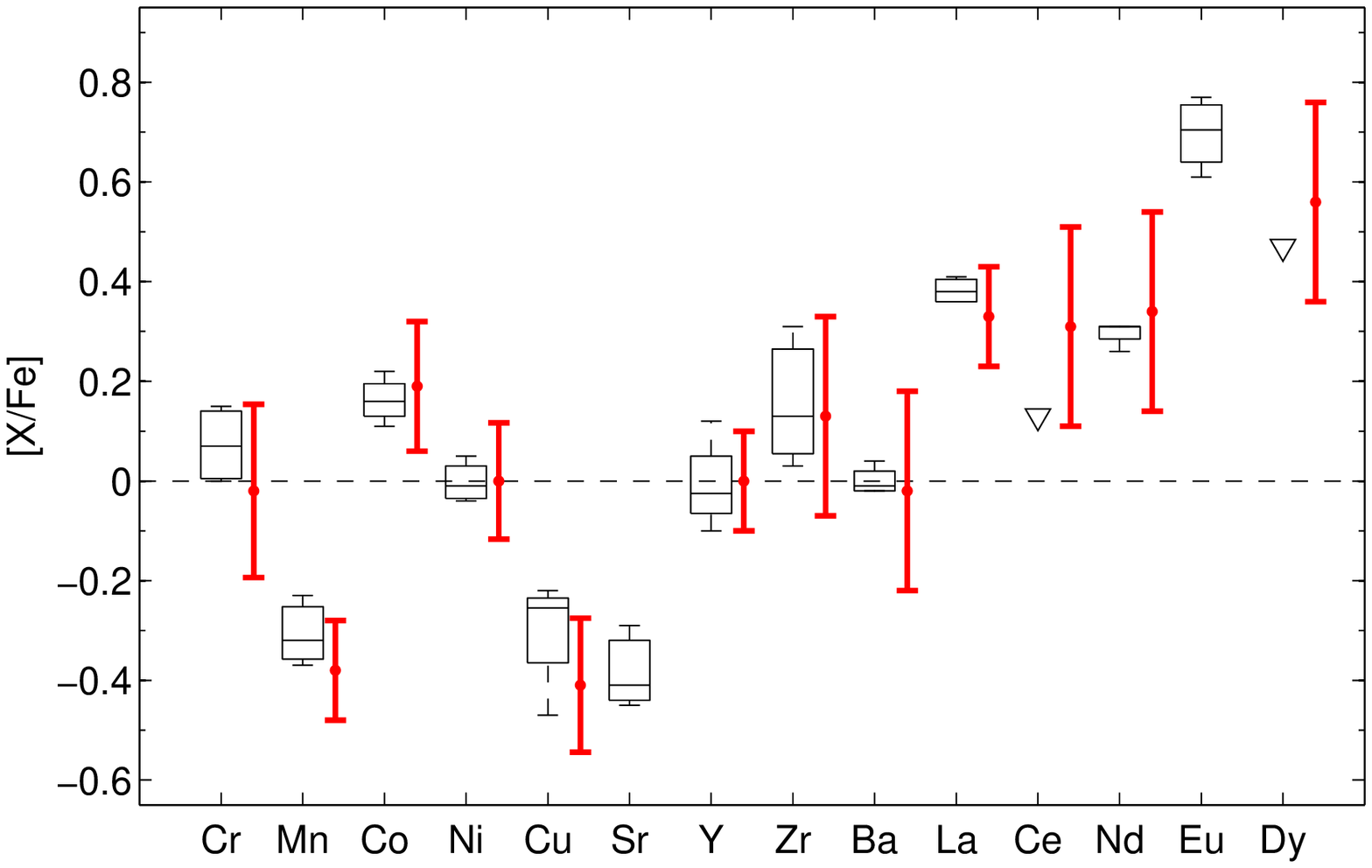}
\caption{Boxplots of Pal~3 abundances from the MIKE spectra, relative to Fe I, for C through V (top) and Cr through Dy (bottom).  
The boxes designate the mean values and interquartile ranges. Iron abundances are shown relative to the cluster mean. The red offset error bars indicate the 
mean and 1$\sigma$ scatter derived from the co-added HIRES spectra, while triangles indicate the elements for which upper limits could 
be derived from co-added MIKE spectra only.}
\end{figure}

In analogy, Table~8 lists the abundance ratios derived from the co-added HIRES spectra. 
\begin{table*}[htb]
\caption{Abundance results from the co-added HIRES spectra}             
\centering          
\begin{tabular}{cccrcccrcccr}     
\hline\hline       
& & All & & & & Pal3-2,3,4,6 & & & & Without Pal3-2,3,4,6 &  \\
\cline{2-4}\cline{6-8}\cline{10-12}
\raisebox{1.5ex}[-1.5ex]{Element} & [X/Fe] & $\sigma$ & N & & [X/Fe] & $\sigma$ & N  & & [X/Fe] & $\sigma$ & N \\
\hline                    
Fe$_{\rm MgI}^a$ & $-$1.58  & 0.18 & \dots & & $-$1.65  & 0.20 & \dots & & $-$1.54  & 0.18 & \dots  \\ 
Fe\,{\sc i}      & $-$1.52  & 0.43 & 113 & & $-$1.49  & 0.69 & 103 & & $-$1.37  & 0.54 & 105  \\
Fe\,{\sc ii}     & $-$1.32  & 0.36 &   5 & & $-$1.27  & 0.48 &   3 & & $-$1.22  & 0.35  &   5  \\
Mg\,{\sc i}      & 0.43  & \dots & 1 & & 0.24  & 0.03 & 3 & & \dots  & \dots & \dots  \\
Al\,{\sc i}      & 0.68  & \dots & 1 & & 0.56  & \dots & 1 & & \dots  & \dots & \dots  \\
Si\,{\sc i}      & 0.34  & 0.28 & 3 & & 0.40  & 0.29 & 4 & & 0.30  & 0.37 & 3  \\
Ca\,{\sc i}      & 0.38 & 0.20 & 13 & & 0.31  & 0.35 & 11 & & 0.29  & 0.20 & 16  \\
Sc\,{\sc ii}     & 0.26  & 0.33 & 4 & & 0.16 & 0.20 &  4 & & 0.22 & 0.14    & 4  \\
Ti\,{\sc i}      & 0.24  & 0.32 & 18 & & 0.13 & 0.18 & 11 & & 0.26  & 0.40 & 13  \\
Ti\,{\sc ii}     & 0.35  & 0.32 & 9 & & 0.31  & 0.46 & 4 & & 0.30  & 0.21 & 6  \\
V\,{\sc i}       & 0.14  & 0.23 & 4 & & 0.15  & 0.22 & 3 & & 0.13  & 0.08 & 3  \\ 
Cr\,{\sc i}      & $-$0.02  & 0.46 & 7 & & $-$0.19  & 0.44 & 7 & & $-$0.03  & 0.37 & 8  \\
Mn\,{\sc i}      & $-$0.38  & 0.02 & 2 & & $-$0.30  & \dots & 1 & & $-$0.44  & \dots & 1  \\
Co\,{\sc i}      & 0.19  & 0.26 & 4 & & 0.04& 0.28 & 4 & & 0.00  & 0.01 & 2  \\
Ni\,{\sc i}      & 0.00  & 0.35 & 9 & & $-$0.13  & 0.18 & 6 & & 0.06  & 0.17& 9  \\
Cu\,{\sc i}      & $-$0.41  & 0.19 & 2 & & $-$0.62  & \dots & 1 & & $-$0.40  & 0.13 & 2  \\
Y\,{\sc ii}       & 0.00  & 0.15 & 3 & & \dots  & \dots & \dots & & 0.00  & 0.20 & 3  \\
Zr\,{\sc ii}     & 0.13  & \dots & 1 & & 0.11  & \dots & 1 & & 0.24  & \dots & 1  \\
Ba\,{\sc ii}     & $-$0.02  & \dots & 1 & & $-$0.03  & \dots & 1 & & $-$0.04  & \dots & 1  \\
La\,{\sc ii}     & 0.33  & 0.10 & 3 & & 0.24  & \dots & 1 & & \dots   & \dots & 1  \\
Ce\,{\sc ii}     & 0.31  & \dots & 1 & & 0.29 & \dots & 1 & & 0.13  & \dots & 1  \\
Nd\,{\sc ii}     & 0.34  & \dots & 1 & & 0.30  & \dots & 1 & & 0.30  & \dots & 1  \\
Dy\,{\sc ii}     & 0.56  & \dots & 1 & & \dots & \dots & \dots & & 0.87  & \dots & 1  \\
\hline                  
\end{tabular}
\begin{flushleft}
$^a$Metallicity estimate based on the Mg I calibration of Walker 
et al. (2007), on the metallicity scale of Carretta \& Gratton (1997).
\end{flushleft}
\end{table*}
These result are broadly consistent with those derived from the individual MIKE spectra to within the uncertainties, as also indicated by the red error bars in Fig.~3,  which depict the mean [X/Fe] and 1$\sigma$ spread obtained from co-adding all HIRES spectra.  
As Fig.~4 shows, the results are invariant against the sample used for co-addition.
In the following plots, we will include the data point from co-adding {\em all} HIRES spectra, but will focus the discussion and statistics 
of the Pal 3 abundances from the more accurate MIKE data. 
\begin{figure}[htb]
\centering
\includegraphics[width=1\hsize]{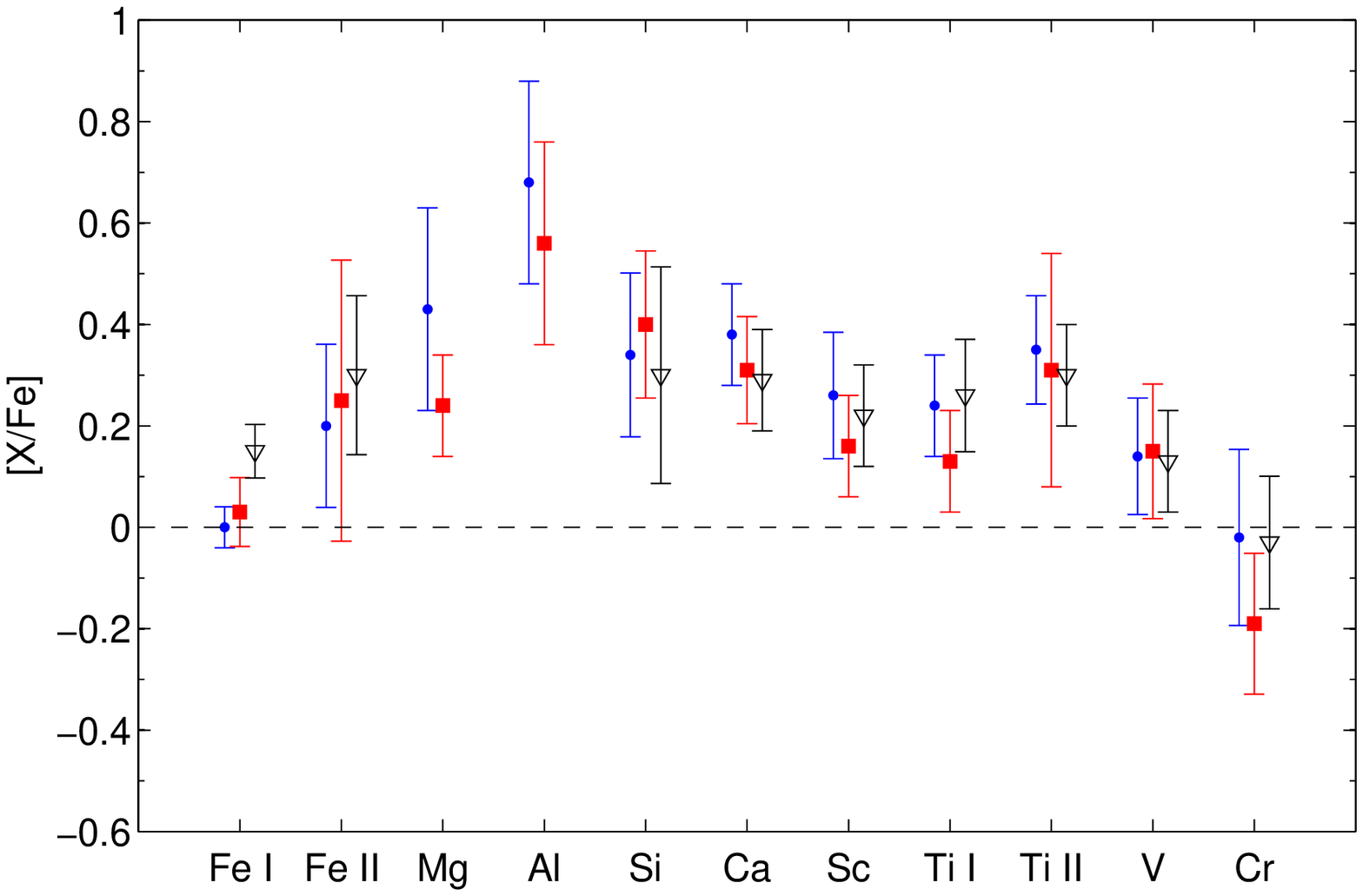}
\includegraphics[width=1\hsize]{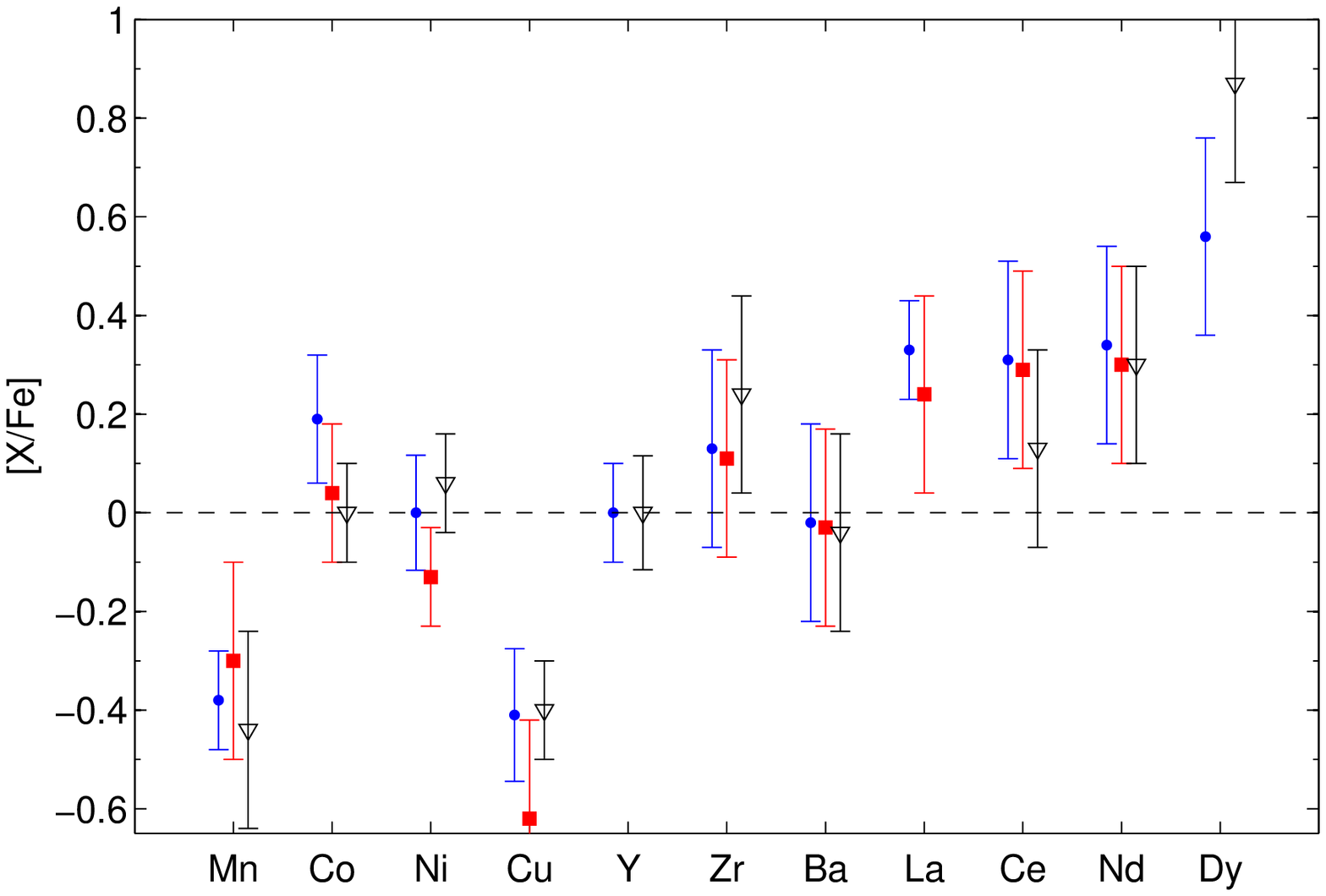}
\caption{Elemental abundances derived from co-added HIRES spectra, for O through Cr (top) and Mn through Dy (bottom), 
including different samples of stars: all HIRES stars (blue circles), Pal3-2,3,4 and 6 only (red  
squares), and all stars but excluding the latter (black triangles).}
\end{figure}

In Fig.~5 we show the run of the [$X$/Fe] abundance ratios with effective temperature exemplary  for the $\alpha$-elements O through Ti.
\begin{figure}[htb]
\centering
\includegraphics[width=1\hsize]{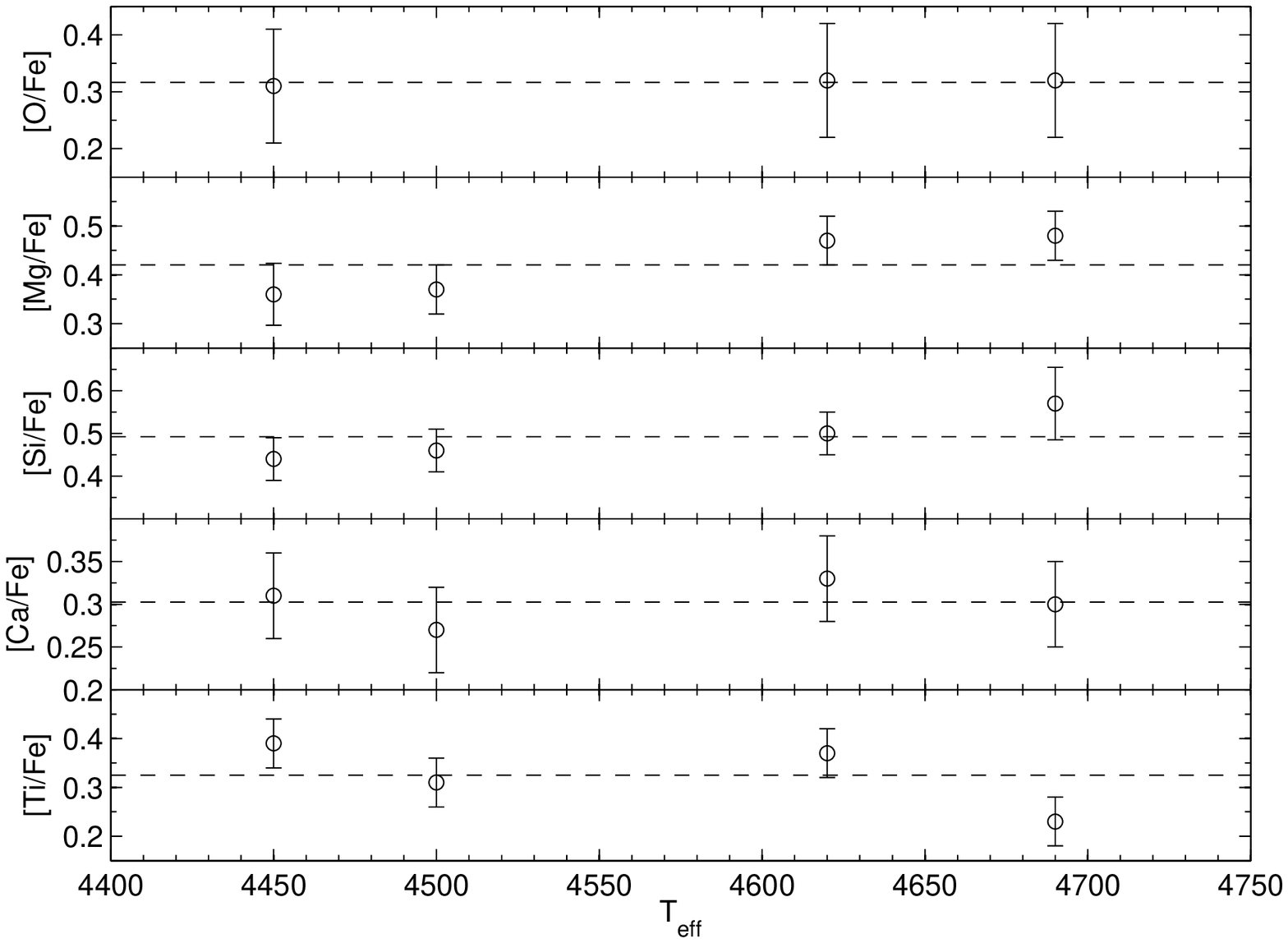}
\caption{Abundance ratios for O, Mg, Si, Ca and Ti as a function of stellar effective temperature. Shown as  dashed lines are the cluster mean values. Error bars are 1$\sigma$ random errors, 
where a minimum error was assumed for elements with only one measurable line.}
\end{figure}
All abundance ratios, with the possible exception of Mg and Si (see also Sections~5.4 and 6.1) 
are consistent with showing no trend with evolutionary stage in that they 
are constant with respect to T$_{\rm eff}$ to within the uncertainties.
We now briefly comment on our findings for the groups of individual elements.   
\subsection{Iron abundance}
From the four red giants with MIKE spectra we find a cluster mean [Fe/H] from the neutral lines of $-1.58\pm0.02$ (stat.) $\pm0.13$ (sys.) dex. 
This is fully consistent with the CaT based mean metallicity of  $-1.60\pm0.02$ dex derived above. 
In particular, we find a mean discrepancy of the CaT and Fe I values of our four stars of only $-$0.02$\pm$0.02 dex. 
In fact, both our CaT estimate and the more accurate [Fe/H] abundance ratio are in excellent agreement with the 
estimates of Stetson et al. (1999) from CMD fitting and the CaT measurement of Armandroff et al. (1998)\footnote{Note, 
however, that these authors placed their measurements on the GC abundance scale of Zinn \& West (1984), which 
yields [Fe/H] larger by up to 0.3 dex in this metallicity regime.}. Our value is also marginally consistent with 
the range of metallicities given by Ortolani \& Gratton (1989) and Hilker (2006) to within the uncertainties. 
Analysis of the co-added HIRES spectra yields a slightly higher value of $-$1.52$\pm$0.04, which nevertheless agrees with the value from the 
accurate MIKE analysis to within the uncertainties.  

We do not find any trend of [Fe/H] with T$_{\rm eff}$, which is a good indicator of stellar evolutionary status. This reassures us that co-adding individual spectra 
will yield abundance results that are representative of the entire cluster (recall from Fig.~2 that the HIRES targets span $\sim$600 K across the RGB). 
Moreover, there is no discernible trend of a deviation of Fe I and II with T$_{\rm eff}$, so that LTE is a valid approximation at the temperatures and metallicities 
of our stars (see also Th\'evenin \& Idiart 1999; Ram\'irez \& Cohen 2003; Yong et al. 2005).

On the other hand, the ionised  iron abundances derived from the co-added HIRES spectra are systematically higher than the neutral 
values, with a 
mean [Fe I/ Fe II] of $-0.19 \pm 0.02$. 
Within the combined error bars, this discrepancy  is significant at 1.4$\sigma$ on average. 
Typical EWs of the ionised  lines range from 30--120 m\AA. 
One might argue that non-LTE effects start to affect our stars more strongly  the further down we move on the RGB, so that 
the co-added spectra suffer from the integrated non-LTE corrections. On the other hand, a large deviation of  $-$0.22 dex is already 
present when only the four brightest spectra (Pal3-2 through Pal3-6) are combined (see middle part of Table~8). 
Moreover, in their integrated light abundance study of 47~Tuc, 
McWilliam \& Bernstein (2008) find a small discrepancy between Fe I and Fe II of 0.03 dex that is well within the respective uncertainties. 
Whatever the cause, we conclude that  Fe II lines seem in general not suitable for establishing a population's iron abundance from a low S/N spectral co-addition as 
employed in the present work (cf. Kraft \& Ivans 2003). 

\subsection{C, O abundances}
Although the low S/N ratios of the individual blue spectra did not allow us to resolve the CH G-band at 4323\AA, we derived a constraint  
of the mean [C/Fe] abundance ratio by co-adding the four MIKE spectra. 
We then fit synthetic spectra with stellar parameters that provided a representative mean  of the actual MIKE targets (Table~2) 
by varying carbon abundances in a least-squares sense to the stacked spectrum.   
For this purpose, we employed a CH line list with isotope splitting, assuming a  $^{12}$C/$^{13}$C ratio of 10, as typically found in evolved giants, 
and $gf$-values by B. Plez (A. Frebel, private communication; see Frebel et al. 2007).  
As Fig.~10 implies, the co-added spectrum is reasonably well fit with a [C/Fe] ratio of $\sim$$-$0.36$\pm$0.25 dex, which we state as an 
order of magnitude estimate of Pal 3's overall carbon abundance. 
\begin{figure}[htb]
\centering
\includegraphics[width=1\hsize]{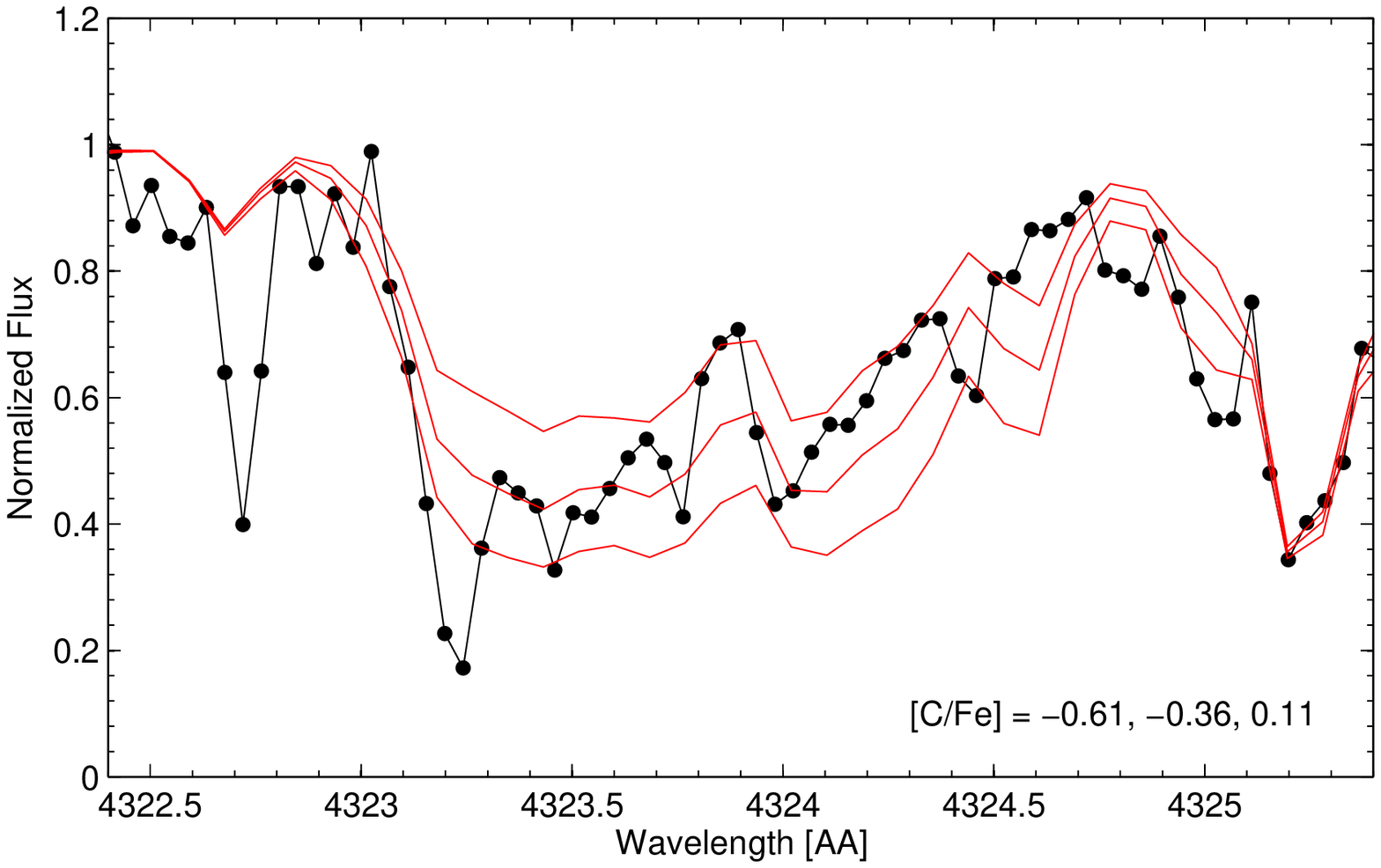}
\caption{Co-added MIKE spectra around the CH feature at 4323\AA. Shown as red lines are the best fit synthetic spectrum and 
those with [C/Fe] differing by $\pm$0.25 dex.}
\end{figure}
Given the  luminosities of our targets (on the upper RGB; $\log\left(L/L_{\odot}\right)\sim2.7$), the [C/Fe] ratio we find 
is fully representative of evolved stars that had typical standard carbon abundances when they formed, and which   
were depleted to the observed level in the course of their stellar evolution  (Gratton et al. 2004; Aoki et al. 2007; see also Frebel et al. 2009; their Fig.~11). 

Oxygen abundances could not be directly inferred from EW measurements of the weak 
[OI] 6300, 6363\AA~lines, as these are heavily affected by sky emission lines. 
Instead, we synthesized the spectral region around the 6300\AA~line and varied the [O/Fe] abundance to achieve an acceptable fit to the  
portion of the stellar line that is unaffected by telluric [O I] in the line wing. 
Since this procedure is still hampered by a low S/N around the absorption feature, we assigned a minimum error bar of 0.2 dex to the  [O/Fe] values. 
 For Pal3-3, the emission residual directly coincides with the stellar line and no [O/Fe] could be derived for this star. 
 As a result, we find a mild [O/Fe] enhancement in our stars of typically 0.3 dex. 
\subsection{Light odd-$Z$ elements: Na, Al, K}
The Na-D 5889, 5895\AA~lines, which are present in our MIKE spectra,  generally yield abundance results that differ systematically from those derived from 
the near-infrared set of lines at  8183, 8194\AA~(e.g., Ivans et al. 2001; Ram\'irez \& Cohen 2003).  
Since the former resonance lines are too strong at the metallicities in question and strongly affected by 
telluric absorption to be measured reliably,  we chose to measure the sodium abundance ratios from the 8193, 8194\AA~doublet instead 
of averaging the full set of the four deviant lines. 
Unfortunately, no Na abundances could be inferred from the HIRES spectra, since the \mbox{Na-D} falls on the spectral gaps and the near-infrared lines are 
not covered by our set-up.  
For stars with similar atmosphere parameters to our Pal~3 stars,  the calculations 
of Takeda et al. (2003) show that downward non-LTE corrections 
become more severe with increasing EW for the Na I lines 
at 8183 and 8195\AA. At those relatively large EWs of $\sim$140m\AA~ in our stars, the Takeda et al. (2003) 
results suggest abundance corrections that reduce our LTE values by 0.45--0.50, if  non-LTE effects on Fe are ignored. 
This would bring the sodium abundances in accord with the slightly depleted Galactic halo stars (Fig.~7). 
Given the complexity of the non-LTE, we proceed by adopting the LTE abundance ratios for further discussions and list those in Table~7. 
\begin{figure}[htb]
\centering
\includegraphics[width=1\hsize]{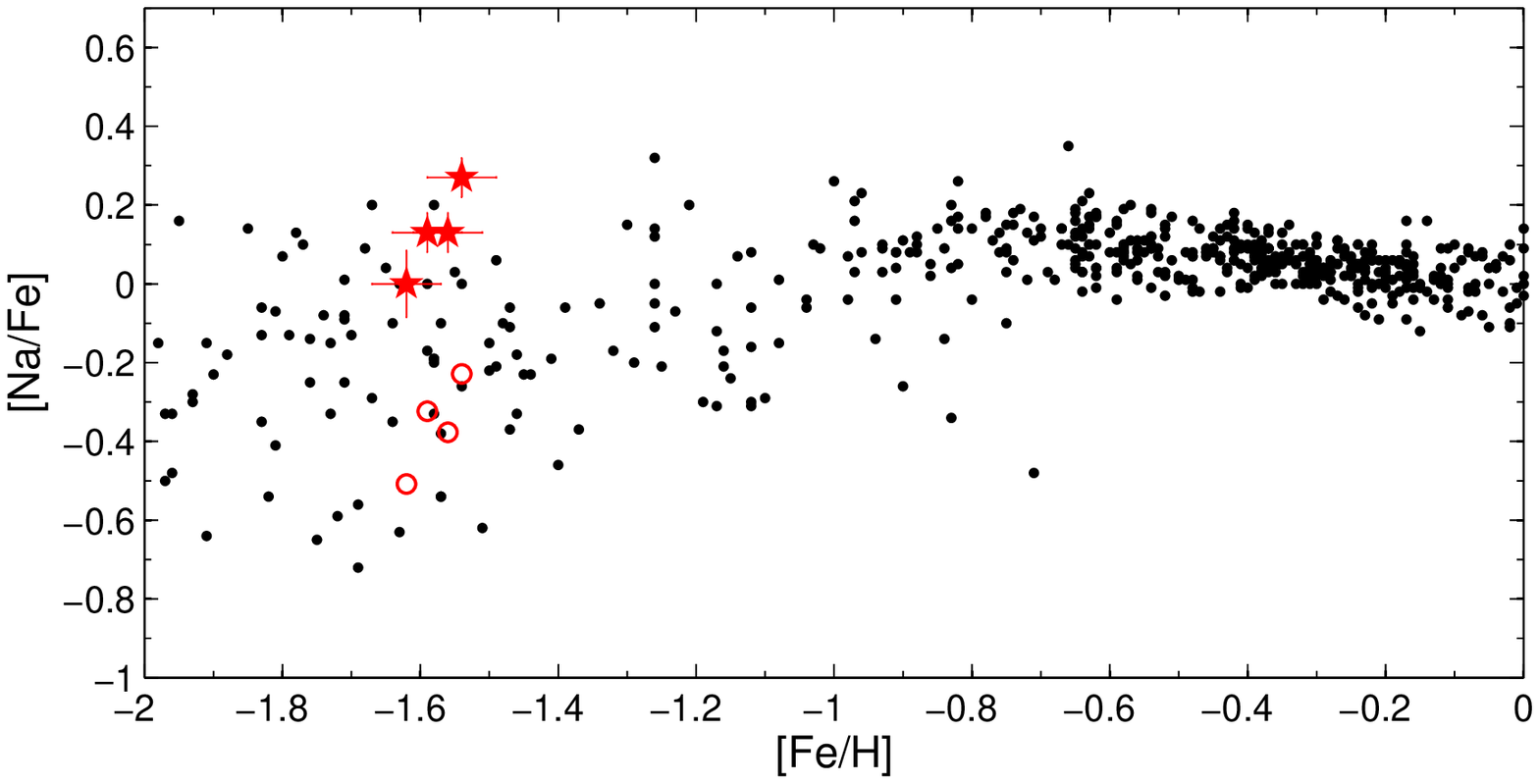}
\caption{Sodium abundance ratios for the Pal 3 stars (red symbols) and the Galactic halo and disks (black dots; see text for references). 
The star symbols show our derived LTE abundance ratios, while open circles illustrate the 
abundances after applying the Na non-LTE corrections from Takeda et al. (2003) for stars at the same 
metallicities as Pal~3.}
\end{figure}
Since we only have [O/Fe] abundances available for three of our stars, it is not fully representative to test our limited sample for any anti-correlations between 
Na and O within Pal 3 itself.  The [O/Fe] ratio in our stars is approximately constant at 0.30 dex, but we note that 
 the nominal [Na/O] ratios of our stars are 
 consistent with the trends outlined by the numerous samples of Galactic GC stars (Fig.~8; Gratton et al. 2004; Carretta 2006). 
The present data can neither confirm nor refute the Na-O anti-correlation in Pal~3.
\begin{figure}
\centering
\includegraphics[width=1\hsize]{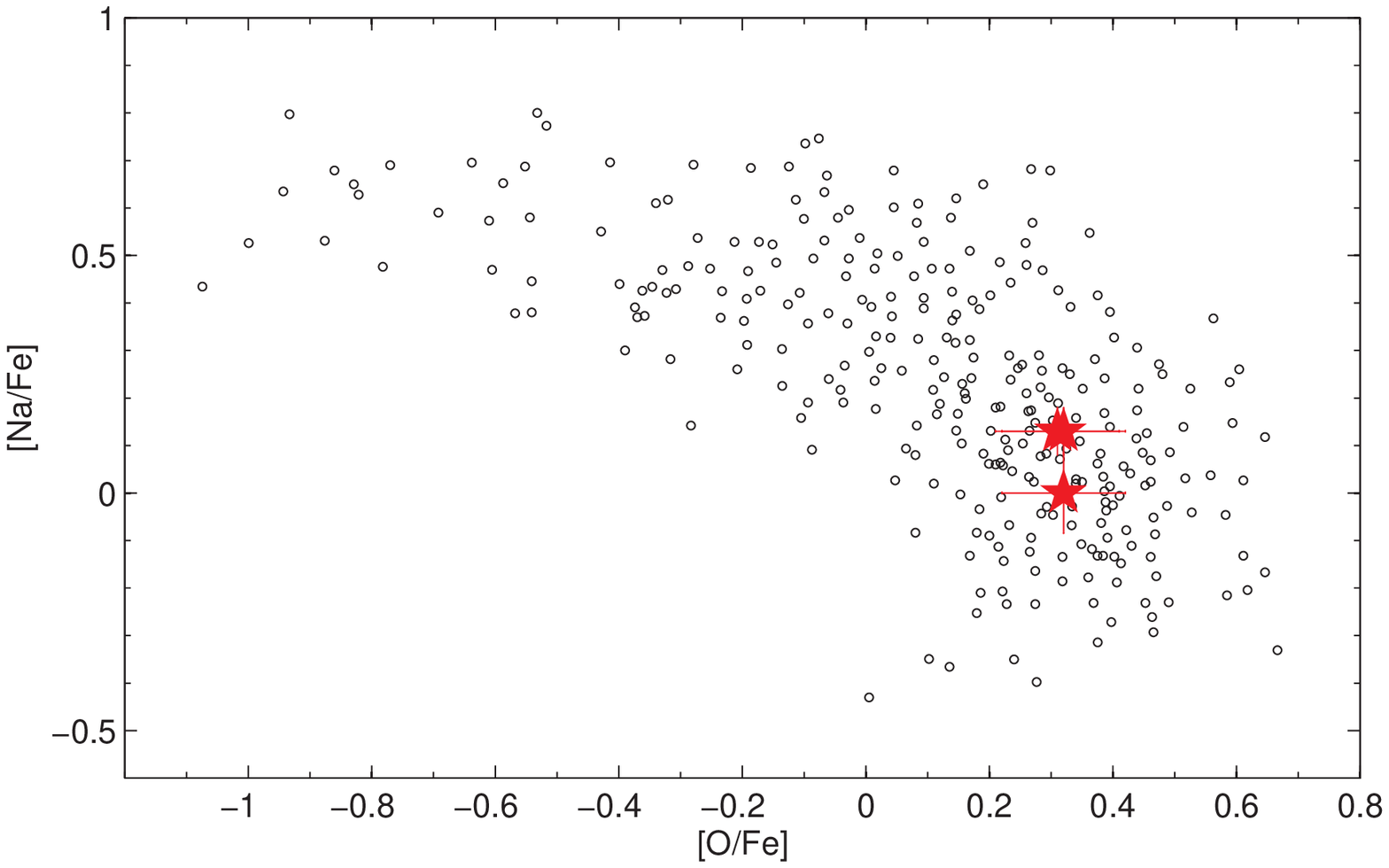}
\caption{Na-O Anti-correlation for  Galactic GC stars (small black circles) after  Gratton et al. (2004) and Carretta (2006).
Our Pal~3 measurements are shown as red stars.}
\end{figure}

Unfortunately, no aluminum line could be detected in any of our MIKE spectra. However, we derived upper limits from marginal detections of the 
6696\AA~line in the co-added HIRES spectra. Thus, we find an [Al/Fe] of $\la 0.7$ dex, which is in accord with the values found in 
GCs of comparable metallicities (e.g., Cohen \& Mel\'endez 2005a). 

The determination of potassium abundances in red giants is a delicate venture for two reasons. First, the two strongest resonance lines at 7698, 7664\AA~ fall into a window of 
strong telluric contamination. Second, the [K/Fe] ratio is strongly affected by non-LTE effects. 
Regarding the first issue, we are fortunate that  the 7698\AA\ resonance line in Pal 3 is sufficiently free from telluric absorption at the GC's radial velocity.   
With respect to the departure from LTE, Takeda et al. (2009) suggest that downward  
non-LTE corrections in mildly metal poor GC giants can be as high as  $\sim -$0.35 dex (see also Zhang et al. 2006). 
At the metallicity of M5 ($\sim$$-$1.5 dex), Takeda et al. (2009) predict NLTE corrections of the order of $-$0.2 dex, albeit for cooler giants than in our sample. 
As for the case of sodium, an accurate treatment of the departure from LTE in our stars is clearly beyond the scope of this paper, but we note that 
an extrapolation of the  Takeda et al. (2009)  corrections to the T$_{\rm eff}$ of our targets, would result in our  K-abundances being lowered by  0.4--0.6 dex. 
\subsection{The $\alpha$-elements: Mg, Si, Ca, Ti}
Mg and Si are the only elements for which we find an indication of increasing abundances with increasing T$_{\rm eff}$ 
of the order of (0.7$\pm$0.3)  dex\,(1000 K)$^{-1}$. 
It should be noted, however, that the Si abundance is generally based on 2--5 weak lines that are often affected by low S/N 
and, at their high excitation potential,  may be prone to temperature uncertainties. 

Likewise, the Ti II abundance in Pal3-3 was derived from only two lines, since the other transitions were too strongly affected by noise or 
blends.
All in all, the [$\alpha$/Fe] abundance ratios of Mg,Si,Ca, and Ti are enhanced to about 0.4 dex, which is the canonical value found in Galactic  halo field stars 
and GCs over a broad range of metallicities (see also Sect.~6; Fig.~10) and Galactocentric radii. 
\subsection{Iron peak elements: Sc, V, Cr, Mn, Co, Ni, Cu}
In all our stars the even-Z iron-peak element ratios  [Cr/Fe] and [Ni/Fe] follow the abundance of iron so that 
[X/Fe]$\sim$0. This trend is representative of the values found in the halo population and other GCs (e.g., Yong et al. 2005). 
Nissen \& Schuster  (1997)  reported a significant correlation of their [Ni/Fe] abundance ratios in Galactic disk and halo stars with [Na/Fe], which 
was subsequently confirmed for other stellar systems such as the dSphs (e.g., Shetrone et al. 2003). 
As for the Na-O anti-correlation, our sample of four stars is too small to investigate any such intrinsic correlation in Pal~3, but 
we note that our values agree well with the overall [Ni/Na] relation defined by Galactic stars. 
 
For Sc, V, and Co, we find ratios that are mildly enhanced to $\sim$0.2 dex and thus slightly larger than the Solar 
 values found in the other clusters (e.g., Yong et al. 2005; Cohen \& Mel\'endez 2005a), which may be an artifact of the usually 
low S/N around the few weak absorption lines in question, while Cr and Ni have more and better defined transitions in our spectra.  
The  odd-Z elements Mn and Cu are depleted with respect to iron; their mean values of $-$0.31 ([Mn/Fe]) and 
$-$0.30 ([Cu/Fe]) are consistent with 
the element ratios found in a number of Galactic GCs and the moderately metal poor halo field 
stars, albeit falling towards the higher end of the envelope  (Cayrel et al. 2004; McWilliam et al. 2003; McWilliam \& Smecker-Hane 2005). 
We note, however, that [Cu/Fe] has been derived from the  5105\AA~line only, which may be affected by blends that lead to  larger 
uncertainties. 
\subsection{Neutron-capture elements: Sr, Y, Zr, Ba, La, Nd, Eu\\(Ce, Dy)}
Although lying in the blue, low-S/N part of our MIKE spectra, we estimated [Sr/Fe] from the 4077, 4215\AA~resonance lines that, with EWs of 200--250m\AA, are 
strong in our stars, yet reach consistent abundances. 
 This yields depleted Sr abundance ratios of typically $-0.3$ dex. 
The abundances of the n-capture elements Y, La and Nd were determined from typically 2--3 sufficiently strong lines, while the 
[Zr/Fe] ratio is solely based on the weak  5112\AA~line. 
[Eu/Fe], on the other hand, was measured from the weak ($\sim$20--30 m\AA) 6437 and 6645\AA~lines. 
As a result, we find slightly supersolar Y and [Zr/Fe] abundance ratios, whilst the elements heavier than La are enhanced 
to [X/Fe] of 0.35 dex up to 0.8 dex for [Eu/Fe], which we also confirmed from spectral synthesis and measurements in a higher S/N co-added MIKE 
spectrum. 
The [Ba/Fe] ratios, derived from the 5853, 6141, 6496\AA~lines, are  found to be Solar and are therefore  compatible 
with the [Ba/Fe] ratios of Galactic halo stars above [Fe/H]$\ga$$-2$ dex. We note, however, that this may be an artifact of the 
high microturbulent velocities adopted for our stars. While most transitions for other elements are relatively unaffected by this 
parameter, the Ba lines are very sensitive to changes in $\xi$. For example, if we were to assume a lower microturbulence of $\sim$1.7 km\,s$^{-1}$ 
as suggested by a canonical T$_{\rm eff}$- or log\,$g$-$\xi$ relation (e.g., Cayrel et al. 2004; Yong et al. 2005), our stars would have 
higher [Ba/Fe] abundance ratios with a mean of 0.42 dex. As the adopted values for $\xi$ yield a consistent EW equilibrium, 
we proceed by using the higher values, but urge caution in the interpretation of our measured [Ba/Fe]. 

Although neither Ce nor Dy could be measured from the blue, low-S/N part of our individual MIKE spectra, we derived upper limits 
from a co-added spectrum of the four stars (as in  Sect.~3.2),  therefore yielding an estimate of the GC's mean abundance ratios 
for these elements. This procedure succeeded for the weak Ce 5274.2\AA~and Dy 5169.8\AA~lines, from which we derived 
a [Ce II /Fe II] ([Dy II /Fe II]) of 0.13 (0.47) dex, respectively. 
The transition probabilities for these were taken from Sadakane et al. (2004; and references therein) and  
we based the error on the resulting [Ce, Dy/Fe] ratios on the acceptable range in measured EWs, which {led} to 
a conservative  estimate of $\sim\pm$0.15 dex. 
In Fig.~9, we compare our observed abundance ratios for the neutron-capture elements (Z$\ge$38) to the Solar-scaled abundance patterns (Burris et al. 2000), 
where we normalised the curves to the same Ba abundance, both for the values obtained for the high microturbulence of our stars (top panel) and 
for a lower value of 1.7 km s$^{-1}$ (bottom panel).   
\begin{figure}
\centering
\includegraphics[width=1\hsize]{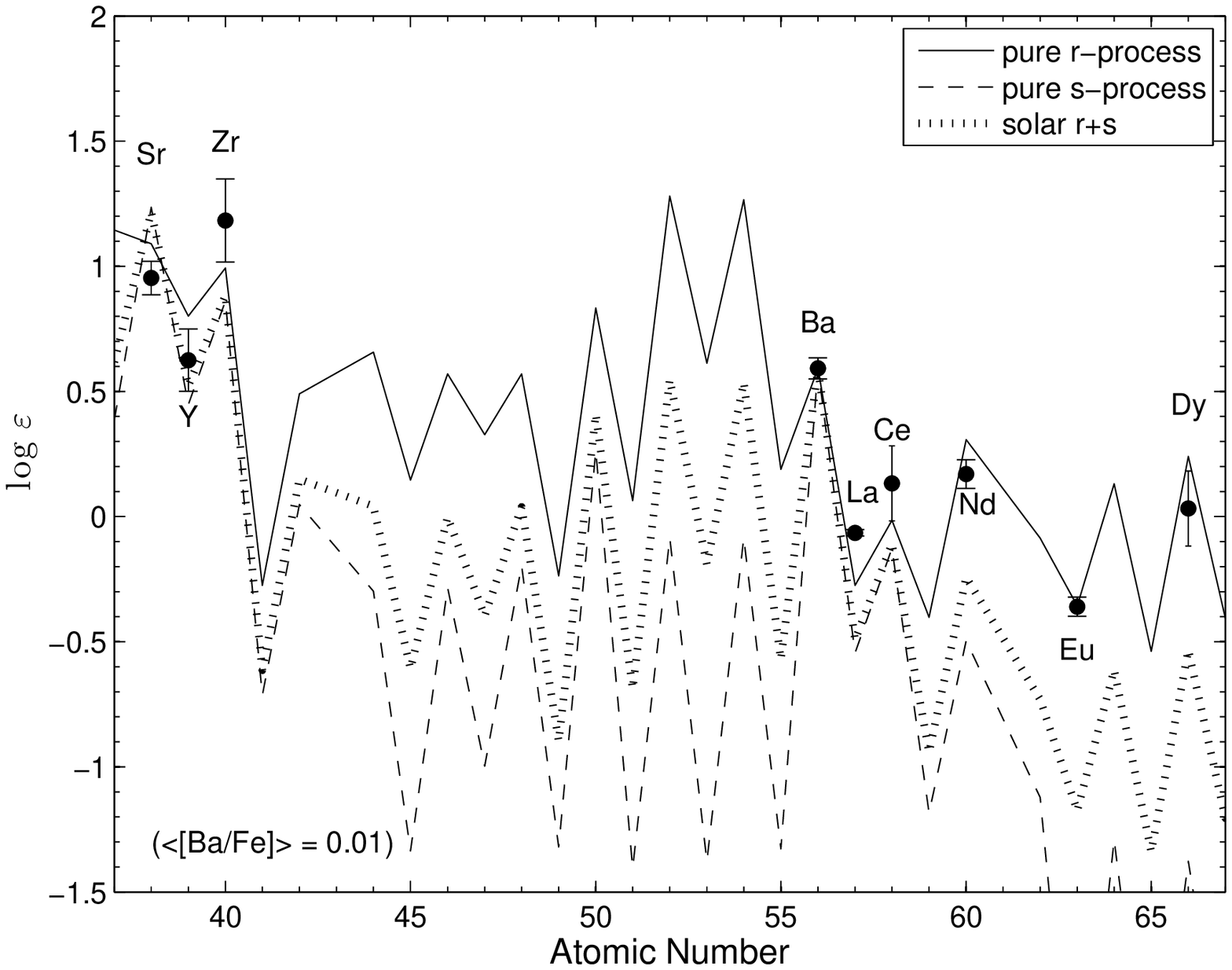}
\includegraphics[width=1\hsize]{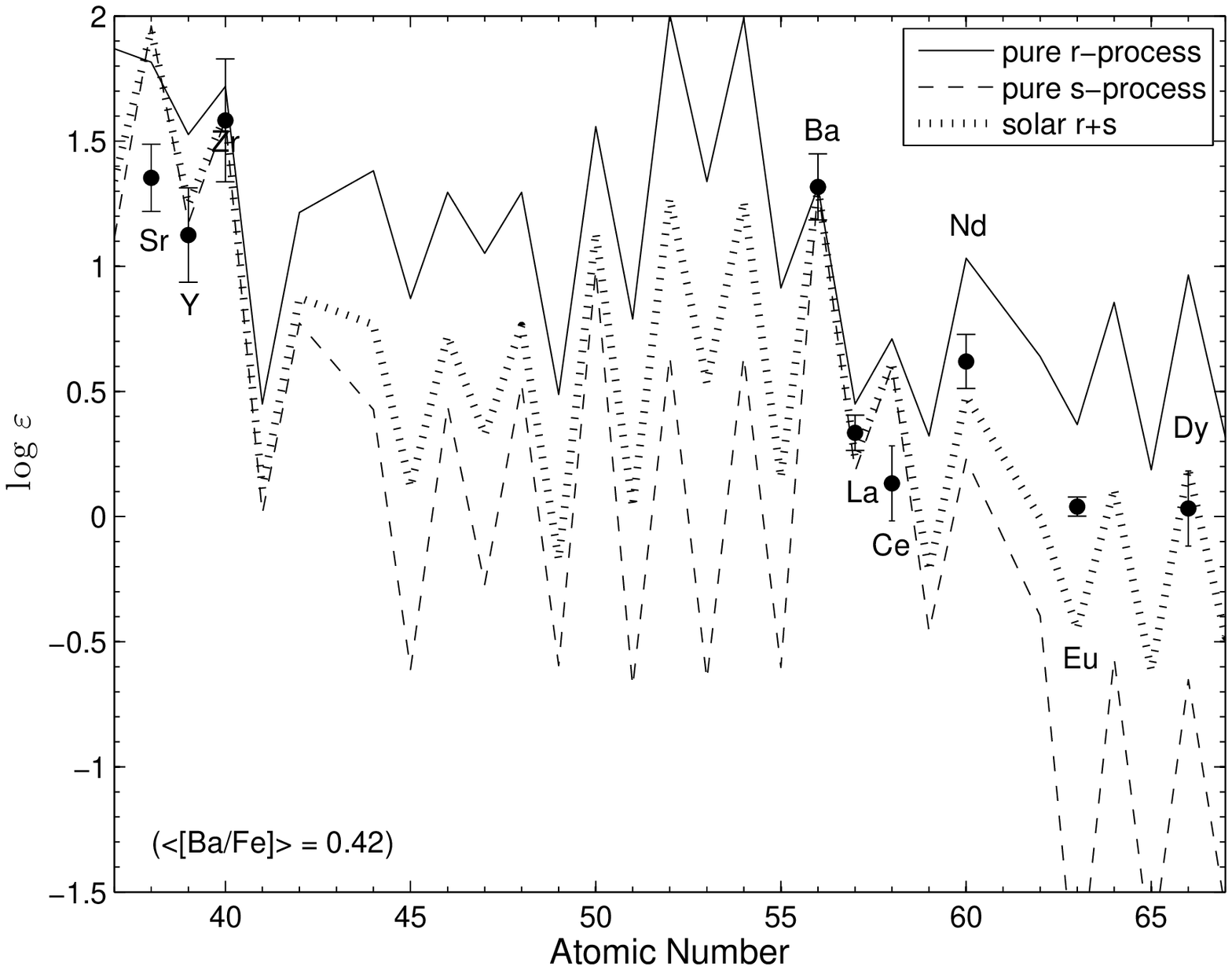}
\caption{Mean neutron-capture abundance ratios in comparison to the scaled-solar r- and s-process ratios of Burris et al. (2000), each normalised to Ba.
The bottom panel assumes a lower microturbulence of $\xi=1.7$ km\,s$^{-1}$ (cf. Table~4), leading to [Ba/Fe] ratios higher by $\sim$0.4 dex. 
The error bars each indicate the $\sigma$-spreads.}
\end{figure}
In the top panel (higher $\xi$), all the heavy elements save Y are fully consistent with the solar $r$-process abundance curve, which implies that these stars formed 
out of material that was not considerably enriched by the AGB stars that produce the $s$-process elements, but 
predominantly by the short-lived supernovae (SNe) of type II.  
Given the only moderately low iron abundance of Pal~3 and the younger age of this cluster, this may seem surprising as the long-lived AGB stars could have 
formed and evolved in the GC's early phases to contribute s-process elements (cf. Sadakane et al. 2004).  If Pal 3's heavy elements are confirmed to be r-process 
dominated, this would be the second known example of a GC with such abundance patterns after M 15 ([Fe/H]=$-$2.3 dex; Sneden et al. 2000). 
Given the possibility of a lower microturbulence, the bottom panel suggests, however, that a higher [Ba/Fe] and therefore a standard $r$+$s$ mixture is not ruled out 
by our data. 
An $r$-process dominance is further underscored by  the relatively 
 high [Eu/Fe] in conjunction with the resulting low [$s$/$r$] ratios at the metallicity of this GC (Fig.~16); note 
 that the unsaturated Eu and La lines are relatively unscathed with respect to the adopted microturbulence. 
 We can further investigate the question of the $r$-process origin of the heavy elements in Pal~3 by differentiating the production channels into 
the {\em weak} $r$-process component (Z$<$56), which may occur in massive SNe~II with progenitor masses above 20 M$_{\odot}$ 
(e.g., Wanajo \& Ishimaru 2006), and the {\em main} $r$-process (all Z$\ge$38; in 8--10  M$_{\odot}$  SNeII; Qian \& Wasserburg 2003). 
Our [Ba/Sr] ratio of 0.35--0.50 dex is  consistent with the values found in halo stars at the same metallicity and is indicative of the 
main $r$-process without any need to invoke an enrichment by very massive stars in Pal~3 as found in low-mass environments such as 
the ultra-faint dSph galaxies (Koch et al. 2008a; Frebel et al. 2009). 
\section{Discussion}
\subsection{(No) Abundance variations} 
As noted above, it is impossible to draw firm conclusions about chemical abundance variations using a low S/N sample of only four stars. 
Nonetheless, to get an order-of-magnitude estimate, we followed Cohen \& Melendez (2005a) in defining the 
spread ratio, SR, as the 1$\sigma$ abundance spread divided by the total systematic and random error (added in quadrature) 
of each element. Thus the SR is an indication of whether the observed spread is a real intrinsic star-to-star scatter or a mere statistical fluctuation. 
In addition, we computed the probability, $P$, that the corresponding (observed)  $\chi^2$ will be exceeded by chance. 
We find SRs ranging from 0.05 to a maximum of 1.05 for Na. These values are close to, or well below, unity so that 
the observed abundances spreads are likely caused by the measurement errors only. 
Furthermore, none of the elements has a $P\,(\chi^2)$ smaller than  0.5\%, which Cohen \& Melendez (2005a) define as a threshold for  
bonafide abundance variations. 

Again, the lowest value is found for Na, with $P\,(\chi^2)=40$\%, while all other elements show no evidence of any variation with 
probabilities much in excess of 60\%. 
While one would expect the scatter of [Na/Fe] to  be accompanied by a comparable spread in [O/Fe], we find no evidence of 
any significant oxygen variation. We note, however, that the SR of 0.05 for [O/Fe] is only based on three stars and derived from crude 
spectral synthesis rather than the EW analysis as for the other elements. 
In this context, the (Kendall rank) probability that Na and O in our stars are uncorrelated within the random errors  is (77$\pm$32)\%, but again 
we note the potential bias of a low number statistics coupled with different measurement methods of these two elements. 
Given the close resemblance of our abundance data with  stars in other GCs (Fig.~8; Sect.~6.2), we conclude 
that Pal~3 probably follows the usual abundance variations and anti-correlations caused by internal mixing 
processes in the giant stars (e.g., Ivans et al. 2001; Gratton et al. 2004),  even if the present, sparse data cannot confirm nor rule out 
any stronger trend.  
Essentially the same result holds for the Na-Ni correlation  (at a probability of 70$\pm$28~\%) found in Galactic field and GC stars, 
with which the Pal~3 abundance ratios are broadly consistent. 

The SRs of [Mg/Fe] and [Si/Fe] are 0.52 and 0.76, which confirms that neither of these elements shows any significant variations. This also suggests 
that their marginal trend with T$_{\rm eff}$ (Fig.~5) is not a real feature. 
The smallest range of abundances is found for the heavy elements Ba, La, and Nd with P\,$(\chi^2)$ of 99.3--99.6\%, which is in accord with  the 
findings in other GCs (e.g., Yong et al. 2005), and for Ca at a comparably high level.  
All in all, we conclude that our limited data are compatible with no star-to-star scatter. A detailed evaluation of the marginal evidence of
the canonical abundance (anti-)~correlations (Na/O, Na/Ni) within Pal 3 must await more data. 
Such a homogeneity in the elements supports the possibility of abundance analyses from co-added spectra of individual stars without the 
introduction of scatter due to star-to-star variations in the final GC spectrum. 

Furthermore, this characteristic clearly rules out Pal~3 being a low-luminosity dwarf galaxy. 
All such systems studied to date (e.g., Shetrone et al. 2001, 2003; Kirby et al. 2008; Koch  2009) show large abundance spreads of 
the order of several tenths of a dex. 

\subsection{Comparison with other GCs}
As mentioned in the previous Sections, the chemical abundance distributions of Pal 3 show few surprises. 
{\em All} of the elements studied in this work are fully consistent with those ratios found in GCs across a wide range of metallicity and 
Galactocentric radii. We illustrate this behavior in Figs.~10--12, in which we show the run of the mean 
[$\alpha$/Fe]\footnote{Although this average of [(Mg+Ca+Ti)/ 3 Fe] is a convenient and illustrative measure, one should keep in mind that 
Mg, Ca, and Ti are produced in different channels of the SNe II event, therefore limiting somewhat this simplistic mean; see discussions in 
Venn et al. (2004) and Koch et al. (2008b).}, [Ni/Fe] and [Ba/Eu], exemplary for the range of abundance ratios analysed in this work. 
\begin{figure}[htb]
\centering
\includegraphics[width=1\hsize]{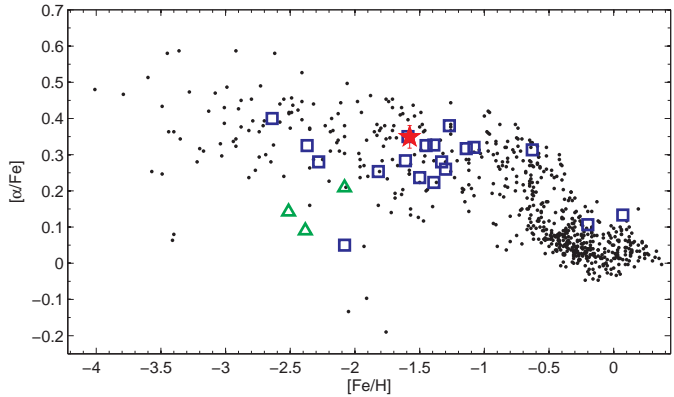}
\caption{Same as Fig.~7, but for the mean [$\alpha$/Fe] abundance ratio. Here, the red star symbol denotes the mean 
and 1$\sigma$-spread of Pal~3  derived from our MIKE spectra. Also 
indicated as green triangles are mean values for the GCs in the Fornax dSph (Letarte et al. 2006). Blue squares are GC data from the literature (see text for references).}
\end{figure}
\begin{figure}[htb]
\centering
\includegraphics[width=1\hsize]{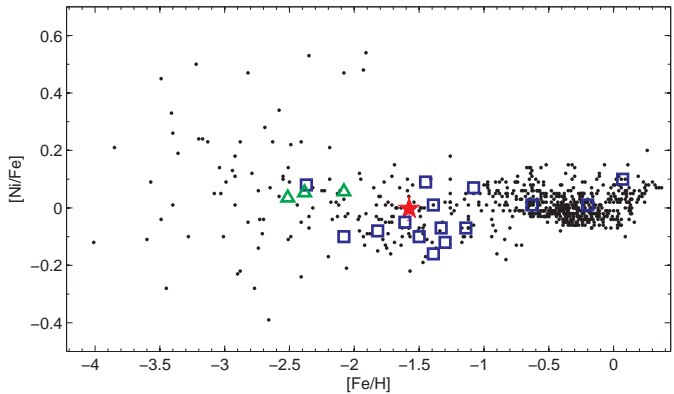}
\caption{Same as Fig.~10, but for the [Ni/Fe] abundance ratio.}
\end{figure}
\begin{figure}[htb]
\centering
\includegraphics[width=1\hsize]{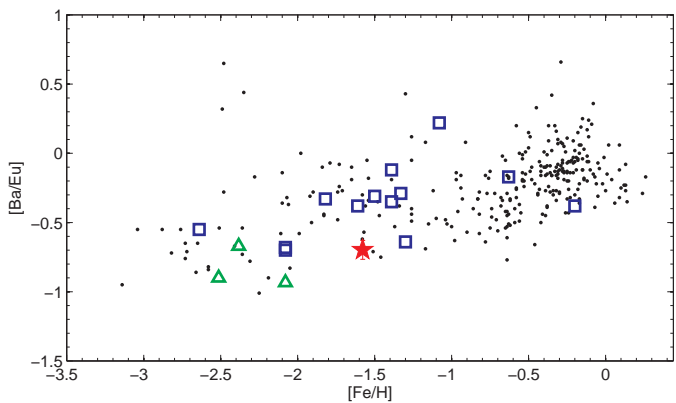}
\caption{Same as Fig.~10, but for the [$s$/$r$] abundance ratio [Ba/Eu].}
\end{figure}
In analogy with Gratton et al. (2004); Pritzl et al. (2005); Cohen \& Mel\'endez (2005a); Yong et al. (2005);  Geisler et al. (2007)  and Koch (2009), 
we overplot in Figs.~7,10,11,12 our measured abundance ratios
on a large number of Galactic thin-disk, thick-disk, and halo stars from the studies of 
Gratton \& Sneden (1988, 1994); Edvardsson (1993);   McWilliam  et al. (1995);  Ryan et al. (1996); Nissen \& Schuster (1997);  McWilliam (1998);  
Hanson et al. (1998); Burris et al. (2000); Prochaska et al. (2000); Fulbright (2000, 2002); Stephens \& Boesgaard (2002); Johnson (2002); 
Bensby et al. (2003); Ivans et al. (2003), and Reddy et al. (2003) (see also Pritzl et al. 2005; Geisler et al. 2007; Koch 2009). 
We further highlight as large open squares a sample of mean GC abundances for 
M~3 (Cohen \& Mel\'endez 2005a);  
M~4 (Ivans et al. 1999);  
M~5 (Ram\'irez \& Cohen 2003);  
M~10 (Haynes et al. 2008);  
M~13 (Cohen \& Mel\'endez 2005a);  
M~15 (Preston et al. 2006);  
M~71 (Ram\'irez \& Cohen 2003);  
M~92 (Shetrone 1996; Sneden et al. 2000); 
NGC~288 (Shetrone \& Keane 2000);  
NGC~362 (Shetrone \& Keane 2000);  
NGC~1851 (Yong \& Grundahl 2008);  
NGC~2419 (Shetrone et al. 2001);  
NGC~2808 (Carretta 2006);  
NGC~3201 (Gonzalez \& Wallerstein 1998);  
NGC~5694 (Lee et al. 2006);  
NGC~6397 (James et al. 2004);  
NGC~6528 (Carretta et al. 2001);  
NGC~6553 (Carretta et al. 2001; Alves-Brito et al. 2006);   
NGC~6752 (Yong et al. 2005);
NGC~7006 (Kraft et al. 1998), and 
NGC~7492 (Cohen \& Mel\'endez 2005b).   
For these plots, no efforts were taken to homogenise the abundance data from the various sources 
with respect to different approaches in 
the analysis (i.e., regarding log~$gf$ values and stellar atmospheres), but we did correct for differences in the adopted Solar abundance scales where necessary 
(see also Cohen \& Mel\'endez 2005b).

As discussed in Sect.~5, the abundance ratios found in Pal~3 are {\em fully compatible with the trends found in the Galactic halo} at comparable metallicities. 
This holds for the individual $\alpha$-elements as well as the iron peak (Fig.~11) and the majority of the heavy, n-capture elements. The [Ba/Eu] ratio 
(Fig.~12) in the Pal~3 stars is slightly lower than in the halo stars and the GCs at [Fe/H]$\sim -1.5$ dex, but it is still consistent with these components 
to within the measured uncertainties. 
We note the following interesting cases: 

{\em (1) NGC 5694}  ([Fe/H]=$-2.08$ dex; Lee et al. 2006): This GC is peculiar because of a strong deficiency in the $\alpha$-elements  
by about 0.3--0.4 dex  compared to the bulk of Galactic GCs\footnote{Note, however, that the respective abundances in this GC were derived 
from a single red giant.}.  Although the heavy element patterns are different from those found in the 
dSph galaxies (e.g., Shetrone et al. 2001, 2003; Geisler et al. 2007; Koch 2009, and references therein), 
such an $\alpha$-depletion, coupled with the large Galactocentric distance of $\sim$30 kpc and its large negative radial velocity, 
prompted Lee et al. (2006) to conclude that this outer halo cluster is likely of extragalactic origin.  
As Fig.~10 indicates,  the  [$\alpha$/Fe] pattern of Pal~3 bears {\em no resemblance with that of NGC~5694}, while their heavy element ratios 
(Figs.~11,12) are comparable to within the uncertainties.  

{\em (2) Fornax GCs}  ($<$[Fe/H]$>$=$-2.33$ dex; Letarte et al. 2006): 
The Fornax dSph is the only MW dSph satellite known to harbour its own GC system apart from the disrupted Sgr system. 
Its abundance ratios are compatible 
with those of the dSph field stars, albeit at fairly low metallicities. In particular, the Fornax GC stars show the canonical 
depletion in [$\alpha$/Fe] with respect to the Galactic halo and are therefore compatible with the ratios of the dSph-like GC NGC~5694 
described above. Their [Ba/Eu] ratios are not unlike the value we measured in Pal~3, but, again, representative of the lower metallicity regime 
around $-2.5$ to $-2.0$ dex. In this comparison, Pal~3 is again dissimilar to the dSph populations. 

{\em (3) NGC 7492}  ([Fe/H]=$-1.82$ dex; Cohen \& Mel\'endez 2005b): Despite its large Galactocentric distance of 25 kpc, this outer halo cluster 
shows abundance ratios that are fully consistent with GCs of the inner halo such as M3 and M13 at around 10 kpc. The intriguing fact that all 
the common chemical elements studied in M~3, M~13 and NGC~7492 are so similar led Cohen \& Mel\'endez (2005b) to conclude that, 
if these clusters were typical representatives of the inner and  outer halos, then these components underwent chemical enrichment
histories that were indistinguishable. 
We therefore compare in Fig.~13 (right panel) our measured abundance ratios to those derived by  Cohen \& Mel\'endez (2005b). 
\begin{figure*}[htb]
\centering
\includegraphics[width=1\hsize]{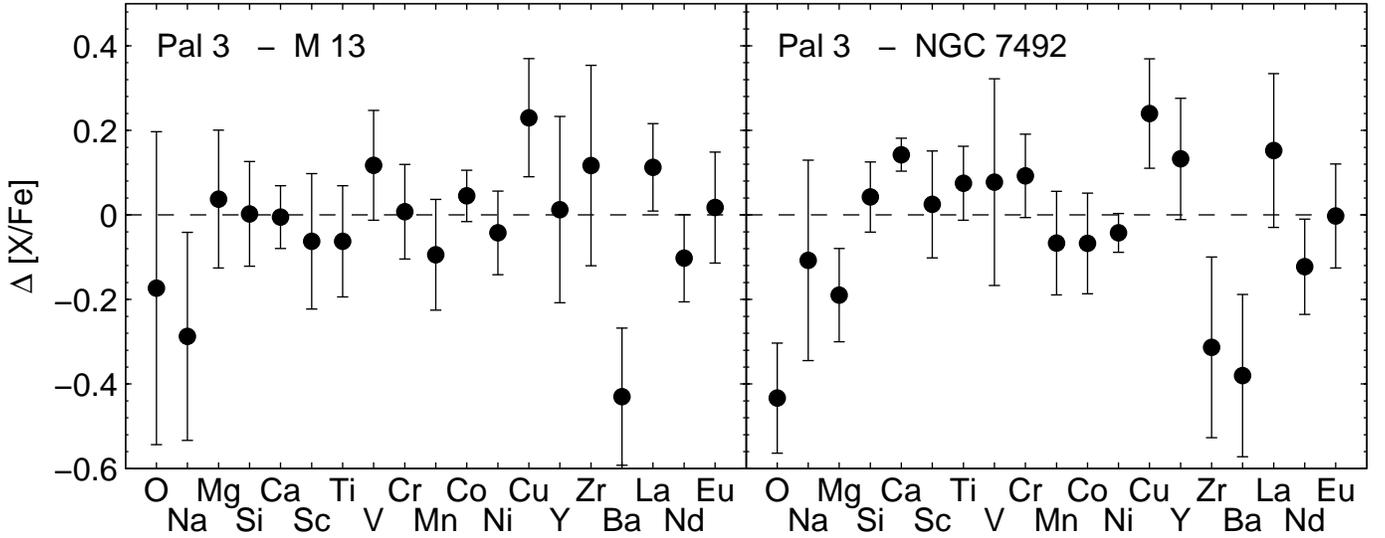}
\caption{Abundance differences in the sense [X/Fe]$_{\rm Pal 3} - $ [X/Fe]$_{\rm GC}$ for the inner halo GC M~13 (R$_{\rm GC}$=12 kpc; [Fe/H]=$-$1.50 dex; 
Cohen \& Mel\'endez 2005a) and the outer halo cluster NGC~7492 (R$_{\rm GC}$=25 kpc; [Fe/H]=$-$1.82 dex; Cohen \& Mel\'endez 2005b). 
We corrected for different Solar abundance scales. Error bars include the 1$\sigma$-spreads for both Pal~3 and the GCs from literature.}
\end{figure*}
The error bars include 1$\sigma$ random errors from both studies, added in quadrature.  To reach a fair comparison, we accounted 
for the different [Fe/H] and Solar abundance scales used in the abundance analyses, but we did not correct for any potential offsets due to variations in 
the line list or stellar atmospheres.  From this we infer that 42\% (63\%) of the abundance ratios that were measured in common between Pal~3 
and NGC~7492 differ by less than 0.10 (0.15) dex. Accounting for measurement uncertainties, this means that 47\% (74\%) of the ratios 
agree to within 1$\sigma$ (2$\sigma$), where the largest discrepancies occur for O, Zr, and Ba. Hence, Pal~3 can probably be considered 
representative of the outer halo GC population in terms of its chemical abundance ratios. 

{\em (4) M3, M13}   ([Fe/H]=$-1.39$, $-1.50$ dex; Cohen \& Mel\'endez 2005a): 
These clusters have long been considered as archetypical GCs and show overall abundance patterns that are in very good agreement with 
other moderately metal poor GCs and Galactic halo field stars, which suggests a common evolutionary history (see  Cohen \& Mel\'endez 2005a 
for a comprehensive discussion). We show in Fig.~17 (left panel) the analogous comparison of the M13 literature abundance distribution 
with our Pal~3 measurements, only correcting for [Fe/H] and the Solar abundances. The mean metallicities of these GCs differ by a mere 0.08 dex 
so that any resemblance in the chemical elements would indicate a common chemical history.  
In fact, 63\% (79\%) of the M13 and Pal~3 abundances measured in common agree to within 0.10 (0.15) dex; that is, 
79\% (95\%) of the chemical elements agree within better than 1$\sigma$ (2$\sigma$). This strengthens the close connection between 
the inner Galactic halo and the outermost regions, as represented by Pal~3. While the co-evolution of the GCs in between $\sim$10 and 
30 kpc (as established by Cohen \& Mel\'endez 2005a) already poses an important constraint on the common history of the
inner and outer halos, the extension of this similarity to the outermost halo at $\sim$100 kpc suggests that the mechanisms of GC formation 
(excluding those that may originate in dSph accretion events) may be invariant over the full extent of the MW halo. 

Following this argument, an $r$-process dominance of the Pal~3 stars would advocate a similar dominance to be found in these inner halo clusters. 
As Fig. 17 implies, this is not the case, since the abundance difference between Pal 3 and either of the GCs is in  fact largest (by $\sim$0.4 dex) 
for [Ba/Fe]. The heavy element and [$s$/$r$] patterns in M3 and M13 found by Cohen \& Mel\'endez (2005a) are compatible with a regular 
Solar $r$+$s$ mix.

\section{Summary}
We have performed a comprehensive abundance analysis of the remote halo GC Pal~3. 
The fact that our co-added high-resolution, low-S/N spectra yield results consistent with individual, higher S/N spectra is an important step 
towards future analyses of faint and remote systems, for which no abundance information can yet be gleaned. 
Although systematic uncertainties and the low S/N ratios complicate such studies, an accuracy of 0.2 dex on most abundance ratios yields 
information sufficient to place such systems in context with both the inner and outer halo GCs, and the faint dSph galaxies. 
We were unable to detect significant abundance variations in this GC, with the possible exception of sodium. Our Pal~3 sample is hampered by 
a small number of stars; within the current data it cannot be ruled out that also this GC follows the global (Na/O, Na/Ni) correlations defined 
by Galactic GC stars (Gratton et al. 2004).
This clearly contrasts the large abundance spreads found in all dSphs studied to date and argues against Pal 3 being an accreted system. 

We find tentative evidence that the heavy elements in Pal 3 are dominated by $r$-process nucleosynthesis, which has to date only been found 
in very metal poor halo field stars and the more metal poor GC M~15. This statement, however, hinges on the adopted value of the microturbulence. 
Further studies from higher S/N spectra are clearly desirable to resolve this issue. If real, this finding would pose strong constraints on the cluster's 
early evolution and pollution phases (e.g, Bekki et al. 2007; Marcolini et al. 2009) and give insight into the relevant gas expulsion time scales 
and the cluster environment (e.g., Baumgardt et al. 2005;  Baumgardt \& Kroupa 2007; Parmentier \& Fritze 2009). 

Stetson et al. (1999) noted that  the ``age difference [between Pal~3, M~3 and M~5]  could be smaller if either [Fe/H] or [$\alpha$/Fe] for the outer halo clusters is 
significantly lower than ... assumed''.  
Their CMD fitting suggested an [Fe/H] of $-$1.57 dex, assuming [$\alpha$/Fe]=+0.3 dex. 
Here we have shown that both the iron and $\alpha$-element abundances have nearly these exact values, rendering it likely that 
Pal~3 is younger by up to $\approx$ 2 Gyr than the inner halo reference clusters of comparable metallicity.  
Despite such an age difference, which led to early notions of the existence of two separate GC populations (e.g., Searle \& Zinn 1978), 
we have found that Pal 3 is remarkably typical in its abundance patterns, which are almost identical to 
those in Galactic halo field stars and other, inner (i.e., R$_{\rm GC}<$15 kpc) {\em and} outer (R$_{\rm GC}>$25 kpc) GCs. 
While many authors have established a metallicity dichotomy between the inner and outer halo field stars, such a strong division 
does not appear to extend to the chemical abundances of the MW GC system. 
In short, while an extragalactic origin of the metal-poor stellar halo, e.g., in systems resembling the ultra-faint dSphs (Simon \& Geha 2007; Frebel et al. 2009; 
Koch 2009)  cannot be excluded  at present, our observations would appear to rule out such an accretion origin for  Pal~3.

\begin{acknowledgements}
We are grateful to Anna Frebel for help with the CH line list.
AK acknowledges support by an STFC postdoctoral fellowship. 
 \end{acknowledgements}

\end{document}